\documentclass[superscriptaddress,reprint, PRL]{revtex4-1}
\usepackage[english]{babel}
\usepackage{amsfonts}
\usepackage{amssymb}
\usepackage{amsmath}
\usepackage{cancel}
\usepackage{colortbl}
\usepackage{graphicx}
\usepackage{hyperref}
\usepackage[abs]{overpic}
\usepackage{picture}
\usepackage{rotating}
\usepackage{siunitx}
\usepackage{soul}				
\usepackage{textcomp}
\usepackage{transparent}
\usepackage{xcolor,varwidth}

\usepackage{epstopdf}
\DeclareSymbolFontAlphabet{\mathbb}{AMSb}
\newcommand{\eqr}[1]{Eq.~\eqref{#1}}
\newcommand{\cit}[1]{Ref.~\cite{#1}}

\newcommand{\kbt}{k_{\rm B}T}
\newcommand{\kb}{k_{\rm B}}
\newcommand{\rhos}{\rho_{s}}

\newcommand{\upd}{\mathrm{d}}

\newcommand{\nn}{\nonumber\\}

\begin{document}

\author{Mathijs Janssen}
\email{mjanssen@is.mpg.de}
\affiliation{Max-Planck-Institut f\"{u}r Intelligente Systeme, Heisenbergstr. 3, 70569 Stuttgart, Germany}
\affiliation{Institut f\"{u}r Theoretische Physik IV, Universit\"{a}t Stuttgart, Pfaffenwaldring 57, 70569 Stuttgart, Germany}
\author{Markus Bier}
\email{bier@is.mpg.de}
\affiliation{Max Planck Institute f\"{u}r Intelligente Systeme, Heisenbergstr. 3, 70569 Stuttgart, Germany}
\affiliation{Institut f\"{u}r Theoretische Physik IV, Universit\"{a}t Stuttgart, Pfaffenwaldring 57, 70569 Stuttgart, Germany}
\affiliation{Fakult\"{a}t Angewandte Natur- und Geisteswissenschaften, Hochschule f\"{u}r Angewandte Wissenschaften W\"{u}rzburg-Schweinfurt, Ignaz-Sch\"{o}n-Str. 11, 97421 Schweinfurt, Germany}

\date{\today}

\begin{abstract}
We study the transient response of an electrolytic cell 
subject to a small, suddenly applied temperature increase at one of its two bounding electrode surfaces. 
An inhomogeneous temperature profile then develops, causing, via the Soret effect, 
ionic rearrangements towards a state of polarized ionic charge density $q$ and local salt density $c$. 
For the case of equal cationic and anionic diffusivities, we derive analytical approximations to $q, c$,
 and the thermovoltage $V_{T}$ for early ($t\ll\tau_{T}$) and late ($t\gg\tau_{T}$) times as compared to the relaxation time $\tau_{T}$ of the temperature. 
We challenge the conventional wisdom that the typically large Lewis number, the ratio $a/D$ of thermal to ionic diffusivities, 
of most liquids implies a quickly reached steady-state temperature profile onto which ions relax slowly. 
Though true for the evolution of $c$, it turns out that $q$ (and $V_{T}$) can respond much faster.
Particularly when the cell is much bigger than the Debye length, 
a significant portion of the transient response of the cell falls in the $t\ll\tau_{T}$ regime, for which our 
 approximated $q$ (corroborated by numerics) exhibits a density wave that has not been discussed before in this context. 
For electrolytes with unequal ionic diffusivities, $V_{T}$ exhibits  a two-step relaxation process, in agreement with 
experimental data of Bonetti \textit{et al.} [\href{\doibase 10.1063/1.4923199}{J. Chem. Phys. \textbf{142}, 244708 (2015)}]. 
\end{abstract}

\title{Transient response of an electrolyte to a thermal quench}

\maketitle

\section{Introduction}
The well-known Soret effect refers to the phenomenon that ions dissolved in a nonisothermal fluid can
show preferential movement along or against thermal gradients, characterized by their heats of transport \cite{agar1989single,wurger2010thermal}. 
Determining these ionic heats of transport, both experimentally \cite{costeseque2004transient, bonetti2011huge} and numerically \cite{roemer2013alkali,di2017computational,di2018thermal} is of
primary importance for all applications involving nonisothermal electrolyte solutions, e.g., in colloid and polymer science. 
When ionic thermodiffusion is impeded, for 
instance, by blocking electrodes, local accumulations of either ionic species can be generated.
Since such accumulations are not necessarily charge neutral, applying a temperature difference across an electrolyte can generate a so-called thermovoltage $V_{T}$.
This thermovoltage, the ionic analog of the Seebeck potential 
in semiconductors,
opens the door to energy scavenging from temperature differences \cite{abraham2013high, zhao2016ionic, al2017thermal, liu2018direct}. 
Since an electric current in an external circuit is only present during the transient build-up of $V_{T}(t)$ \cite{wang2017ionic},
it is of interest to study how electrolytic cells respond shortly after a temperature difference is imposed.
Bonetti \textit{et al.} \cite{bonetti2015thermoelectric} experimentally 
found that, after a seemingly instantaneous rise, $V_{T}(t)$ develops with the ``slow'' diffusion
timescale $L^2/D_{+}$, with $2L$ being the electrode separation and $D_{+}$ the cationic diffusion constant.

Theoretical models were developed by 
Agar and Turner \cite{agar1960thermal} and later by Stout and Khair \cite{stout2017diffuse},
who both considered electrolytes with equal cationic and anionic diffusivities, $D_{+}=D_{-}\equiv D$.
Motivated by the typically large ratio ${a}/D\approx 100$, with $a$ being the thermal
diffusivity, 
their analyses departed from the ansatz
that the steady-state temperature profile develops instantaneously [$T(x,t)=T(x)$, 
with $x$ being the spatial coordinate of their one-dimensional model electrolytic cells],
after which ions relax slowly. 
With this ansatz, 
an exact expression for the transient response of the neutral salt density $c(x,t)$ \cite{agar1960thermal} 
and approximate expressions for the
ionic charge density $q(x,t)$ and corresponding
$V_{T}(t)$ \cite{stout2017diffuse} were found. 
As we show in the present article, the corresponding exact solutions to $q$ and $V_{T}$ 
decay at late times with a common timescale
$\tau_{q}=L^2/(D[(\kappa L)^2+\pi^2/4])$, with $\kappa$ being the inverse Debye length. 
This timescale and particularly the appearance of $\kappa$ therein presents us with two major problems. 
 The first problem is that $\tau_{q}\approx1/(D\kappa^2)$ for large systems ($\kappa L\gg1$), which does not explain the experimental observations of  
\cit{bonetti2015thermoelectric} who found that $V_{T}$ develops much slower.
As we show in this article, this discrepancy does not arise when one accounts for unequal diffusivities among ions.
The second, conceptual, problem that $\tau_{q}$ hints at is  
that the ansatz $T(x,t)=T(x)$ can be unjustified. 
To see this, consider the ratio of the timescales of the
pure thermal relaxation of the cell in the absence of ions [timescale $\tau_{T}=4L^2/(\pi^2 {a})$, 
c.f. \eqr{eq:pureTrelaxation}]
to that of the ionic charge relaxation:
\begin{equation}\label{eq:ratiotimescales}
\frac{\tau_{T}}{\tau_{q}}= \frac{D}{{a}}\left[1+\frac{4(\kappa L)^{2}}{\pi^{2}}\right] \,.
\end{equation}
Since $\kappa$ depends 
on the salt concentration,  
this ratio can be varied 
over many decades, and is by no means restricted to $\tau_{T}/\tau_{q}\ll1$ 
(requiring minute devices and very low salt concentrations). 
Hence, an instantaneous steady-state temperature profile onto which ions rearrange slowly
is a special case of a more general problem. 
 Given the longstanding experimental and theoretical interest in 
 thermodiffusion of electrolytes \cite{agar1960thermal,stout2017diffuse, costeseque2004transient, bonetti2015thermoelectric, roemer2013alkali,di2017computational,di2018thermal, bonetti2011huge,abraham2013high}, 
 it is timely to discuss its solution.

\section{Setup}\label{sec:setup}
\begin{figure}
\includegraphics[width=8.6cm]{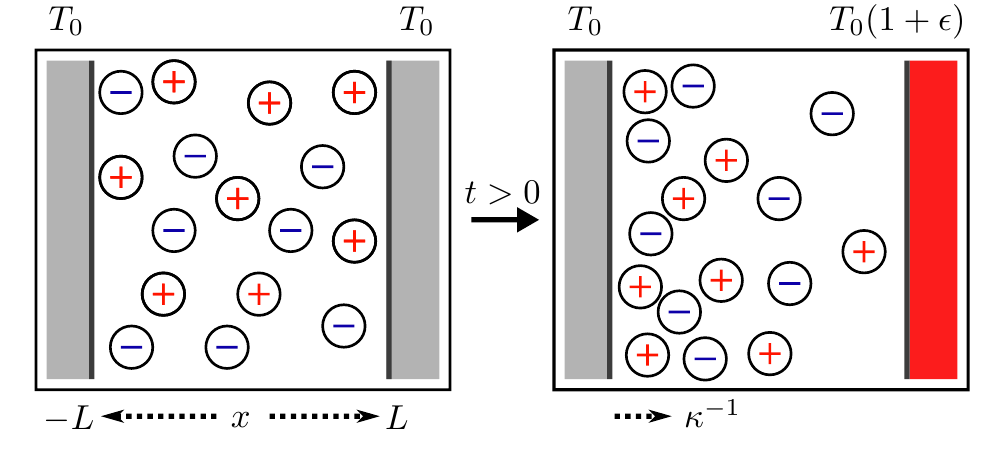}
\caption{A model thermoelectric cell consisting of a 1:1 electrolyte with Debye length $\kappa^{-1}$ (solvent not shown) and two flat electrodes separated over a distance $2L$. At time $t=0$, 
the temperature of one electrode increases by a factor $1+\epsilon$.}
\label{fig1}
\end{figure}

We consider an electrolytic cell (see Fig.~\ref{fig1}) with two parallel flat electrodes at $x=\pm L$. 
Provided that the electrodes are much larger than their
 separation, we can ignore edge effects and treat this system as being one-dimensional.
The electrodes are chemically inert and impermeable to ions and they are not connected by an external circuit, hence, do not acquire a surface charge.
The cell is filled with an electrolyte solution at bulk salt concentration $\rhos$ 
in a solvent of dielectric constant $\varepsilon$. 
The valence $z_{i}$ of ionic species $i=\{+,-\}$ is $z_{+}=1$ for the cations 
and $z_{-}=-1$ for the anions, respectively.
The electrolyte is further characterized by ionic diffusion constants $D_{i}$, single-ion heats of transport $Q^{*}_{i}$, 
 the mass density $\varrho$ (kg m$^{-3}$), the specific heat capacity $c_{p}$ (J K$^{-1}$ kg$^{-1}$), 
and the thermal conductivity $\kappa_{\theta}$ (J s$^{-1}$ m$^{-1}$ K$^{-1}$). 
For simplicity, we ignore all (salt) density dependence of these parameters. 
Moreover, we ignore convection here, which is reasonable if temperature differences are small and 
if the thermal gradient is aligned in the direction opposite to gravity \cite{kalaydin2017thermoelectrokinetic}. Alternatively, convection can be minimized in
``microgravity,'' e.g., onboard the International Space Station \cite{triller2018thermodiffusion}.

\subsection{Governing equations}
The electrostatic potential $\psi(x,t)$, the local ionic number densities $\rho_{\pm}(x,t)$,
and the local temperature $T(x,t)$
are modeled via the classical Poisson-Nernst-Planck and heat equations, 
\begin{subequations}\label{eq:electrokinetic}
\begin{align}
\varepsilon_{0}\varepsilon \partial^{2}_{x} \psi&=- e q \,,\label{eq:Poisson}\\
\partial_{t} \rho_{i}&=-\partial_{x} J_{i}\,,\label{eq:continuity_ions}\\
J_{i}&=-D_{i}\left(\partial_{x} \rho_{i}+\frac{z_{i}e\rho_{i}}{\kbt}\partial_{x} \psi +\frac{\rho_{i}Q^{*}_{i}}{\kbt^2}\partial_{x} T\right)\label{eq:flux_ions}\,,\\
\partial_{t} T&={a}\partial_{x}^{2}T-\frac{e}{\varrho c_{p}}(J_{+}-J_{-})\partial_{x} \psi\,,\label{eq:heatequation}
\end{align}
\end{subequations}
with $e$ being the proton charge and $\varepsilon_{0}$ being the vacuum permittivity. 
First, in the Poisson equation [\eqref{eq:Poisson}] appears the ionic unit charge density $q=\rho_{+}-\rho_{-}$.
Next, the Nernst-Planck equations [\eqref{eq:flux_ions}] account for diffusion, electromigration, and thermodiffusion.
Finally, in the heat equation [\eqref{eq:heatequation}] appears ${a}=\kappa_{\theta}/(\varrho c_{p})$, the thermal diffusivity, 
and a heat source term that was discussed at length in Refs.~\cite{janssen2017reversible, janssen2017coulometry}.

Initially, the ionic density profiles and temperature are homogeneous: 
 \begin{align}
  \rho_{i}(x,t<0)&=\rhos\,, &  T(x,t<0)&=T_{0}\label{eq:thermalinit}\,.
 \end{align}
Thereafter, at $t=0$, the temperature of the electrode at $x=L$ is 
suddenly increased to $T(L,t=0)=T_{0}+\Delta T$, with $\Delta T>0$. 
For $t\ge0$, the boundary conditions at the charge-neutral, ion-impermeable electrodes read
\begin{subequations}\label{eqr:bcs}
 \begin{align}
\partial_{x}\psi(\pm L,t)&=0 \,, &  J_{i}(\pm L,t)&=0\,,\label{eq:nofluxbc}\\
T(-L,t)&=T_{0}\,, &  T(L,t)&=T_{0}+\Delta T\label{eq:thermalbc1}\,.
 \end{align}
\end{subequations}
We note that \eqr{eq:nofluxbc} only fixes $\psi$ up to a constant. 
Without loss of generality, we therefore moreover impose 
\begin{align}\label{eqr:bcs2}
\psi(L,t)=0\,.  
\end{align}
This means that the thermovoltage,
 $V_{T}(t)=\psi(-L,t)-\psi(L,t)$, a key observable of our model system, 
simply reads  $V_{T}(t)=\psi(-L,t)$.

\subsection{Dimensionless formulation}\label{sec:dimless}
We nondimensionalize Eqs.~\eqref{eq:electrokinetic}-\eqref{eqr:bcs2} with 
$\tilde{\psi}=\beta_{0}e\psi$ [with $\beta_{0}=1/(\kbt_{0})$], $\tilde{T}=T/T_{0}$, $\tilde{x}=x/L$, 
$\tilde{t}=tD_{+}/L^2$, $\tilde{\rho}_{i}=\rho_{i}/\rhos $, $\tilde{q}=q/\rhos $, $\tilde{J}_{i}=J_{i}L/(D_{+}\rhos)$, and $\tilde{J}_{q}=\tilde{J}_{+}-\tilde{J}_{-}$ to find
\begin{subequations}\label{eq:dimensionlessequations}
\begin{align}
2\partial_{\tilde{x}}^{2}\tilde{\psi}&=- n^2 \tilde{q},\label{eq:Poissondimless}\\
\partial_{\tilde{t}} \tilde{\rho}_{+}&=\partial_{\tilde{x}}\left(\partial_{\tilde{x}} \tilde{\rho}_{+}+
\frac{\tilde{\rho}_{+}}{\tilde{T}}\partial_{\tilde{x}} \tilde{\psi} +2\alpha_{+}
\tilde{\rho}_{+}\partial_{\tilde{x}} \ln\tilde{T}\right)\,,\label{eq:NPdimless}\\
\xi \partial_{\tilde{t}} \tilde{\rho}_{-}&=\partial_{\tilde{x}}\left(\partial_{\tilde{x}} \tilde{\rho}_{-}-
\frac{\tilde{\rho}_{-}}{\tilde{T}}\partial_{\tilde{x}} \tilde{\psi} +2\alpha_{-}
\tilde{\rho}_{-}\partial_{\tilde{x}} \ln\tilde{T}\right)\,,\label{eq:NPdimless2}\\
\partial_{\tilde{t}} \tilde{T}&=\frac{{a}}{D_{+}}\partial_{\tilde{x}}^{2}\tilde{T}-f\tilde{J}_{q}\partial_{\tilde{x}}\tilde{\psi}\label{eq:heatdimless}\,,
\end{align}
\end{subequations}
and
\begin{subequations}\label{eq:dimlessinitandcond}
 \begin{align}
 \tilde{\rho}_{i}(\tilde{x},\tilde{t}<0)&=1\,, & \tilde{T}(\tilde{x},\tilde{t}<0)&=1 \,,\\
\partial_{\tilde{x}}\tilde{\psi}(\pm 1,\tilde{t}\,)&=0 \,,& \tilde{J}_{i}(\pm 1,\tilde{t}\,)&=0\,,\label{eq:nofluxdimless}\\
\tilde{T}(-1,\tilde{t}\ge0)&=1\,,	&\quad
\tilde{T}(1,\tilde{t}\ge0)&=1+\epsilon \,,\label{eq:bcTdimless}\\
\tilde{\psi}(1,\tilde{t}\,)&=0 \label{eq:bcfloatingpsi}\,,
 \end{align}
\end{subequations}
 where $\xi=D_{+}/D_{-}$ represents the ratio of ionic diffusivities, $f=\kb\rhos/(\varrho c_{p})$ is the ionic heat source coupling, 
$\alpha_{i}=Q^{*}_{i}/(2\kbt)$ are the reduced Soret coefficients, and $\epsilon=\Delta T/T_{0}$ measures the size of the thermal quench.
Moreover, $n=\kappa L$ is the dimensionless Debye separation parameter, with $\kappa^{-1}=[\varepsilon_{0}\varepsilon\kbt_{0}/(2\rhos e^2)]^{1/2}$ being the Debye length.
At steady state, $n$ measures to which extent nonzero $q$ values penetrate the bulk: 
While  $n\gg1$ indicates that $q$ is nonzero only in a small region close to the electrode surfaces, if $n\ll1$, the ionic charge imbalance permeates the complete cell [cf. \eqr{eq:exactchargedensity}].
For reasons explained in Sec.~\ref{sec:epsilon}, we omitted $\partial_{\tilde{x}}D_{i}\partial_{\tilde{x}}\tilde{\rho}_{i}$ terms in Eqs.~\eqref{eq:NPdimless} and \eqref{eq:NPdimless2}.

We see that our system is fully specified by seven dimensionless parameters, six of which ($n, \alpha_{\pm}, \xi, {a}/D_{+}$, and  $f$)
appear in \eqr{eq:dimensionlessequations}, and one of which ($\epsilon$) appears in \eqr{eq:bcTdimless}.
In what follows, we will use the dimensionless formulation when we present simplifications to \eqr{eq:dimensionlessequations} 
because this simplifies calculations and because this highlights the roles played by these seven dimensionless parameters. 
However, when we present results, we prefer to restore to conventional units 
because that makes physical interpretation easier.

The next two sections [\ref{sec:Analytical_Approximations} and \ref{sec:Results}] deal with the case $\xi=1$, 
which is a reasonable simplification for several alkali halides. (For,
e.g., KCl, RbBr, CsBr, and RbI, we find $\xi=0.97, 1.00, 1.00$, and $1.01$, respectively \cite{agar1989single}).
We discuss the more general case of $\xi\neq 1$ in Sec.~\ref{sec:unequalD}. 
 Importantly, we will find that the four quantities of our interest ($T$, $\psi$, $q$, and $c$) 
all relax at late times with one of three fundamental timescales: 
the ``thermal diffusion time'' $L^2/a$, the ``diffusion time'' $L^2/D$,  
or the ``Debye time'' $1/(D \kappa^2)$.

\section{Analytical Approximations}\label{sec:Analytical_Approximations}
We aim at deriving analytical approximations to Eqs.~\eqref{eq:dimensionlessequations} and \eqref{eq:dimlessinitandcond} for the case $\xi=1$. 
To do so, we employ an essential simplification of \eqr{eq:dimensionlessequations}, 
namely, that $f\ll1$ for most electrolytes.
For small thermal quenches (cf.~Sec.~\ref{sec:epsilon}) 
this means that the thermal problem [\eqr{eq:heatdimless}] decouples from the ionic problem 
[Eqs.~\eqref{eq:Poissondimless}, \eqref{eq:NPdimless}, and \eqref{eq:NPdimless2}]. 
Accordingly, we first review the thermal relaxation of a pure solvent (Sec.~\ref{Sec:purethermalrelaxation}), 
which serves as input to determine  $q, \psi$, and the local salt density $c=\rho_{+}+\rho_{-}$ (Secs.~\ref{sec:earlytime} and \ref{sec:latetime}).

\subsection{Pure thermal relaxation}\label{Sec:purethermalrelaxation}
In absence of ions, or when the source term of the heat equation is negligible, 
transient thermal response to a boundary value quench is 
governed by a simplified heat equation, 
$\partial_{t} T=a\partial_{x}^{2}T$, 
and the same initial and boundary conditions as in Eqs.~\eqref{eq:thermalinit} and \eqref{eq:thermalbc1}. 
Writing $\mathcal{K}_{j}=j\pi/2$  for $j=1,2,3,\dots$, the solution to this textbook  problem reads \cite{carslaw1959conduction, cole2010heat}
\begin{align}\label{eq:pureTrelaxation}
\frac{T(x,t)-T_{0}}{\Delta T}&=\sum_{j\ge1}\frac{\sin [\mathcal{K}_{j}(x/L-1)]}{\mathcal{K}_{j}}\exp\!\left[-\mathcal{K}_{j}^{2}\frac{at}{L^2}\right]\nn
&\quad+\frac{1+x/L}{2}\,.
\end{align}
The infinite modes of $T$ decay at increasingly short timescales $L^2/(\mathcal{K}_{j}^2{a})$ with increasing $j$:
 the slowest mode ($j=1$) decays with $\tau_{T}\equiv4L^2/(\pi^2 {a})$, i.e., proportional to the thermal diffusion time. 
 
At early times ($t\ll\tau_T$), $T(x,t)$ is barely affected by the Dirichlet boundary condition $T(-L,t)=T_{0}$.
The temperature in the finite-sized cell can then also be modeled by the same heat equation in a semi-infinite geometry $x\in(-\infty,L]$. In that case we have \cite{cole2010heat} 
\begin{align}\label{eq:pureTrelaxationearlyt}
\frac{T(x,t)-T_{0}}{\Delta T}&\approx\textrm{Erfc}\left[\frac{L-x}{2\sqrt{at}}\right] \,.
\end{align}
Naturally, the largest error made with this approximation occurs at the $x=-L$ boundary: $[T(-L,t)-T_{0}]/\Delta T=\textrm{Erfc}\left[L/\sqrt{at}\,\right]=\{2.1\times10^{-45}, 7.8\times10^{-6}, 0.16\}$ 
at $ta/L^2=\{10^{-2}, 10^{-1}, 1\}$, respectively. Hence, \eqr{eq:pureTrelaxationearlyt} can be safely used up to $ta/L^2=10^{-1}$.

\subsection{Small-$\epsilon$ expansions}\label{sec:epsilon}
As we show next, for $\epsilon\ll1$, we can analytically solve Eqs.~\eqref{eq:Poissondimless}, \eqref{eq:NPdimless},  and \eqref{eq:NPdimless2} both at early times [using \eqr{eq:pureTrelaxationearlyt}] and  at late times [using the steady-state limit of \eqr{eq:pureTrelaxation}].
To do so, we expand $\psi$, $q$, and $c$ in the small parameter $\epsilon$:
  $\psi=\psi_{0}+\epsilon \psi_{1} +\mathcal{O}(\epsilon^{2})$,
   $q=q_{0}+\epsilon q_{1}+\mathcal{O}(\epsilon^{2})$, and
     $c=c_{0}+\epsilon c_{1}+\mathcal{O}(\epsilon^{2})$, respectively, and do the same for the remaining five dimensionless parameters:
     $\alpha_{i}=\alpha_{i,0}+\epsilon \alpha_{i,1}+\mathcal{O}(\epsilon^{2})$, etc.
Inserting those variables and parameters into Eqs.~\eqref{eq:Poissondimless}, \eqref{eq:NPdimless}, and \eqref{eq:NPdimless2} results in
$\mathcal{O}(1)$ problems that characterize the initial isothermal situation (clearly, 
$\psi_{0}=0$, $q_{0}=0$, and $c_{0}=2\rhos$), 
and different $\mathcal{O}(\epsilon)$ 
problems for $\psi_{1} , q_{1}$, and $c_{1}$ for the early- and late-time response. 
With a slight abuse of notation, from hereon, we drop the subscript zeros of all dimensionless parameters,
because subscript-one parameters only appear in $\mathcal{O}(\epsilon^{2})$ terms.
Likewise,  if $a$ depends on $T$, Eqs.~\eqref{eq:pureTrelaxation} and \eqref{eq:pureTrelaxationearlyt} apply only if $\epsilon\ll1$.
(When we presented these equations for arbitrary $\epsilon$, we tacitly assumed that $a(T)=a$). 
We moreover note that the source term in  \eqr{eq:heatequation} is $\mathcal{O}(\epsilon^2)$. 
This means that the results of Sec.~\ref{Sec:purethermalrelaxation}, derived by setting $f=0$, 
are accurate for finite $f$ as well. 
Finally, we omitted $\partial_{\tilde{x}}D_{i}\partial_{\tilde{x}}\tilde{\rho}_{i}$ in Eqs.~\eqref{eq:NPdimless} and \eqref{eq:NPdimless2} because these terms are $O(\epsilon^2)$ as well.

\subsection{Early-time ($t\ll\tau_T$) ionic response}\label{sec:earlytime}
Inserting \eqr{eq:pureTrelaxationearlyt} into Eqs.~\eqref{eq:NPdimless} and \eqref{eq:NPdimless2} yields
\begin{subequations}\label{eq:NPearlyt}
\begin{align}
\partial_{\tilde{t}}\tilde{q}_{1}&=\partial^{2}_{\tilde{x}}\tilde{q}_{1}-n^2 \tilde{q}_{1}+\frac{{\alpha_{\rm d}}(1-\tilde{x})}{\sqrt{\pi}(a\tilde{t}/D)^{3/2}}\exp{\bigg[-\frac{D(1-\tilde{x})^2}{4a\tilde{t}}\bigg]}\label{eq:NPearlyqt}\\
\partial_{\tilde{t}}\tilde{c}_{1}&=\partial^{2}_{\tilde{x}}\tilde{c}_{1}+\frac{{\alpha_{\rm s}}(1-\tilde{x})}{\sqrt{\pi}(a\tilde{t}/D)^{3/2}}\exp{\bigg[-\frac{D(1-\tilde{x})^2}{4a\tilde{t}}\bigg]}\label{eq:NPearlyct}\,,
\end{align}
\end{subequations}
with  $\alpha_{\rm d}=\alpha_{+}-\alpha_{-}$ and $\alpha_{\rm s}=\alpha_{+}+\alpha_{-}$. 
The $n^2\tilde{q}_{1}$ term in \eqr{eq:NPearlyqt} stems from the electromotive term $2\partial^{2}_{\tilde{x}}\tilde{\psi}_{1}$ in $\tilde{J}_{q}$,
together with \eqr{eq:Poissondimless}.
The corresponding  electromotive term in the salt flow 
$\partial_{\tilde{x}}[\tilde{q}_{1}\partial_{\tilde{x}}\tilde{\psi}_{1}]$ is $\mathcal{O}(\epsilon^{2})$ thus neglected in \eqr{eq:NPearlyct}.

At time $t=0$, the system is charge neutral [$q(x,t)=0$] and the nonzero ionic charge current is caused solely by thermodiffusion.
There will be early (but finite) times at which thermodiffusion still dominates electromigration: times, thus, 
 at which the electromotive term $n^2\tilde{q}_{1}$ in \eqr{eq:NPearlyqt}
can be neglected. Clearly, (1) we cannot expect to find a self-consistent nonzero solution for $q(x,t)$ in this way and (2) 
the temporal range of validity of this approximation will decrease with increasing $n$.
With the omission of the $n^2\tilde{q}_{1}$ term  in \eqr{eq:NPearlyqt}, the 
equations governing the  early-time response of $\tilde{q}_{1}/\alpha_{d}$ and $\tilde{c}_{1}/\alpha_{\rm s}$ are the same.
Since the same equations have the same solutions, our forthcoming results for $q_{1}$ are trivially transferable to $c_{1}$.
Substituting $p=D(\tilde{x}-1)^2/(4a\tilde{t}\,)$ in \eqr{eq:NPearlyqt} yields a inhomogeneous ordinary differential equation:
\begin{subequations}\label{eq:diffeqp}
\begin{align}
p\frac{\upd^{2}\tilde{q}_{1}}{\upd^{}p^2}+\left(\frac{1}{2}+p\frac{a}{D}\right)\frac{\upd^{}\tilde{q}_{1}}{\upd^{}p}&=-\frac{2{\alpha_{d}}}{\sqrt{\pi}}\sqrt{p}\exp{[-p]}\label{eq:diffeqp1}\,,\\
\sqrt{p}\frac{\upd^{}\tilde{q}_{1}}{\upd^{}p}\bigg|_{p=0}&=\frac{2{\alpha_{d}}}{\sqrt{\pi}}\,,\label{eq:diffeqpbc1}
\end{align}
\end{subequations}
where \eqr{eq:diffeqpbc1} follows from $\tilde{J}_{q}(1,\tilde{t}\,)=0$  [cf. \eqr{eq:nofluxdimless}].
While \eqr{eq:diffeqpbc1} fixes one of the two integration constants of the general solution of \eqr{eq:diffeqp1},
it turns out that the other integration constant cannot be fixed by  $\tilde{J}_{q}(\tilde{x}=-1,t)=0$; 
we simply do not have the freedom to impose $\upd \tilde{q}_{1}/\upd p$ in two positions.
This must be because \eqr{eq:diffeqp1} resulted from a procedure that ignores the electrode at $\tilde{x}=-1$. 
To fix this second integration constant nevertheless, we enforce charge neutrality 
$\int_{-1}^{1}\upd\tilde{x} \,\tilde{q}_{1}(\tilde{x},\tilde{t}\,)=0\Rightarrow\int_{0}^{D/(a\tilde{t}\,)}\upd p\,\tilde{q}_{1}(p)/\sqrt{p}=0$,
which arises naturally from Eqs.~\eqref{eq:continuity_ions} and \eqref{eq:nofluxbc}.
We find
\begin{align}\label{eq:earlyq}
\frac{q_{1}(x,t)}{2\rhos \alpha_{d}}
&=\frac{2D}{D-a}\Bigg\{\frac{\sqrt{at}}{L\sqrt{\pi}}\left(\exp\left[-\frac{L^2}{Dt}\right]-\exp\left[-\frac{L^2}{at}\right]\right)\nn
&\quad\quad\,\,\,-\sqrt{\frac{{a}}{{D}}}\textrm{Erf}\left[\frac{L-x}{2\sqrt{Dt}}\right]+\textrm{Erf}\left[\frac{L-x}{2\sqrt{a t}}\right]\nn
&\quad\quad\,\,+\sqrt{\frac{{a}}{{D}}}\textrm{Erf}\left[\frac{L}{\sqrt{Dt}}\right]-\textrm{Erf}\left[\frac{L}{\sqrt{a t}}\right]\Bigg\}\,,
\end{align}
and the same for  $c_{1}/(2\rhos \alpha_{s})$.

We can now find $\psi(x,t)$ by integrating $q_{1}$ twice and enforcing 
$\psi(L,t)=0$  and $\partial_{x}\psi(L,t)=0$. The solution, which is  too lengthy to be reproduced here,
turns out to satisfy the boundary condition $\partial_{x}\psi(-L,t)=0$ as well. 
This means that the electromotive term drops out of $\tilde{J}_{q}(-1,\tilde{t})$. We determined the importance of the two remaining
terms in $\tilde{J}_{q}(-1,\tilde{t})=-\partial_{\tilde{x}}\tilde{q}_{1}(-1,\tilde{t})-2\alpha_{d}\partial_{\tilde{x}}\ln \tilde{T}(-1,\tilde{t})$ 
with Eqs.~\eqref{eq:pureTrelaxationearlyt} and \eqref{eq:earlyq}
and found for $a/D=100$
that
$\tilde{J}_{q}(-1,\tilde{t})=\{10^{-43}, 1.8\times10^{-4}\}$ at $ta/L^2=\{10^{-1},1\}$, respectively.
Hence, as long as \eqr{eq:pureTrelaxationearlyt} approximates $T(x,t)$ decently,
 the boundary condition $\tilde{J}_{q}(-1,\tilde{t}\,)=0$ that we could not strictly impose is satisfied approximately
nevertheless.

\subsection{Late-time ($t\gg\tau_T$) ionic response}\label{sec:latetime}
Upon inserting the steady-state temperature profile
$\tilde{T}=1+\epsilon (1+\tilde{x})/2$, at $\mathcal{O}(\epsilon)$, 
Eqs.~\eqref{eq:NPdimless} and  \eqref{eq:NPdimless} give rise to
\begin{align}\label{eq:electrokineticslowions}
\partial_{\tilde{t}} \tilde{q}_{1}&=\partial^{2}_{\tilde{x}} \tilde{q}_{1}-n^{2}\tilde{q}_{1}\,,
&\quad\quad\quad
\partial_{\tilde{t}} \tilde{c}_{1}&=\partial^{2}_{\tilde{x}} \tilde{c}_{1}\,.
\end{align}
Here, the thermodiffusion terms in the ionic fluxes  
amount to constants $\tilde{J}_{i}\sim\epsilon\alpha_{i}$; hence, their spatial derivatives are absent in \eqr{eq:electrokineticslowions}.
As pointed out by Refs.~\cite{costeseque2004transient,stout2017diffuse}, $\alpha_{\rm s}$ and $\alpha_{\rm d}$ then only appear in the boundary conditions, 
\begin{subequations}\label{eq:electrokineticbcslowions}
 \begin{align}
  \tilde{q}_{1}(\tilde{x},\tilde{t}<0)&=0 \,,&  \tilde{c}_{1}(\tilde{x},\tilde{t}<0)&=0 \,,\\
  \partial_{\tilde{x}}\tilde{q}_{1}(\pm 1,\tilde{t}\,)&=-\alpha_{\rm d}\,,&\quad
  \partial_{\tilde{x}}\tilde{c}_{1}(\pm 1,\tilde{t}\,)&=-\alpha_{\rm s} \,,
 \end{align}
\end{subequations}
hence do not affect the relaxation rates.
Only few of the original seven dimensionless numbers controlling Eqs.~\eqref{eq:dimensionlessequations} and \eqref{eq:dimlessinitandcond} now remain. 
We set $\xi=1$ and, by using the steady-state temperature profile, we have effectively set ${a}/D\to\infty$.
With these choices, $\alpha_{\pm}$ moved from the 
PDEs to the BCs. Moreover, as long as $f\lessapprox1$, the source term of the heat equation \eqref{eq:heatdimless} is $\mathcal{O}(\epsilon^2)$ hence irrelevant.
With $\epsilon$ only appearing in the small-$\epsilon$ expansions, $n$ is the 
only remaining parameter that can influence the relaxation rates of our system. 
In Appendix~\ref{sec:exactinversion} we solve Eqs. \eqref{eq:Poissondimless} and \eqref{eq:electrokineticslowions} subject to \eqr{eq:electrokineticbcslowions}.
Writing $\mathcal{N}_{j}=(j-1/2)\pi$ for $j=1,2,3,\dots$, the solutions read
\begin{subequations}\label{eq:exactsolution}
 \begin{align}
\frac{e\psi_{1}(x,t)}{\kbt_{0}\alpha_{\rm d}}&=-n^2\sum_{j\ge1}\frac{1+(-1)^{j}\sin\!\left[\mathcal{N}_{j}x/L\right]}
{\mathcal{N}_{j}^2\left[n^2+\mathcal{N}_{j}^2\right]}\exp{\left[-t/\tau_{q}^{j}\,\right]}\nn
&\quad+\frac{1}{2n}\frac{\sinh (n x/L)-\sinh n}{\cosh  n }+\frac{1}{2}-\frac{x}{2L}\,,\label{eq:exactpotential}\\
\frac{q_{1}(x,t)}{2\rhos \alpha_{\rm d}}&=-2\sum_{j\ge1}\frac{(-1)^{j}\sin\!\left[\mathcal{N}_{j}x/L\right]}
{ n ^2+\mathcal{N}_{j}^2}\exp{\left[-t/\tau_{q}^{j}\,\right]}\nn
&\quad-\frac{1}{n}\frac{\sinh (n x/L)}{\cosh  n }\,,\label{eq:exactchargedensity}\\
\frac{c_{1}(x,t)}{2\rhos\alpha_{\rm s}}&=-2\sum_{j\ge1}\frac{(-1)^{j}\sin\!\left[\mathcal{N}_{j}x/L\right]}
{\mathcal{N}_{j}^2}\exp{\left[-\frac{\mathcal{N}_{j}^2\, Dt}{L^2}\,\right]}\nn
&\quad-\frac{x}{L}\,,\label{eq:exactconcentration}
\end{align}
\end{subequations}
where
 $\tau_{q}^{j}=L^2/[D\left(n ^{2}+\mathcal{N}_{j}^2\right)]$ and where \eqr{eq:exactconcentration} appeared previously in \cit{agar1960thermal}.
We see that, indeed, the relaxation of $\psi$ and $q$ (in units of $L^2/D$) depends only on $n$, while the 
relaxation of $c$ (in units of $L^2/D$) has no parametric dependence whatsoever.
At late times, the relaxation of the functions in \eqr{eq:exactsolution} is dominated by the $j=1$ terms of the sums: 
While $c$ decays with $4L^2/(\pi^2 D)$, $\psi$ and $q$ relax with
$\tau_{q}\equiv L^2/\left[D\left(n^2+\pi^2/4\right)\right]$, as anticipated in the introduction. 
Hence, for $n\ll1$ we find a universal decay time $4L^2/(\pi^2 D)$ proportional to the diffusion time, whereas
for $n\gg1$, $\psi$ and $q$ relax with the Debye time $1/(D\kappa^2)$.

\section{Results}\label{sec:Results}
We numerically solved \eqr{eq:dimensionlessequations} with \textsc{comsol multiphysics} 5.4
for $\xi=1$, ${a}/D=100$, $\alpha_{+}=0.5$, $\alpha_{-}=0.1$, $f=2\times10^{-3}$, $\epsilon=10^{-3}$, and $n=1$ and $n=100$.
This parameter set is representative for an aqueous KCl solution 
($\varrho\approx10^6$ g m$^{-3}$, $c_{p}\approx4$ J g$^{-1}$, $a/D$, and $\alpha_{\pm}$ from \cit{stout2017diffuse}) 
subject to a thermal quench of \mbox{$0.3$ K} around room temperature. 
Because $\kappa^{-1}$ amounts to several tens of nanometers at most (at 1 mM, $\kappa^{-1} = 9.6$ nm), large $n$ values can 
be easily achieved experimentally by using $L$ values in the micrometer regime (or larger). Conversely, the small $n=1$ value 
requires both high dilution and minute devices ($L$ in the nanometer regime). While this latter case might be difficult to reach experimentally, 
we discuss $n=1$ because, judging from \eqr{eq:ratiotimescales}, if anywhere, this is the parameter 
setting for which the instantaneous temperature ansatz [and \eqr{eq:exactsolution}] should work best.

\begin{figure}
\includegraphics[width=8.6cm]{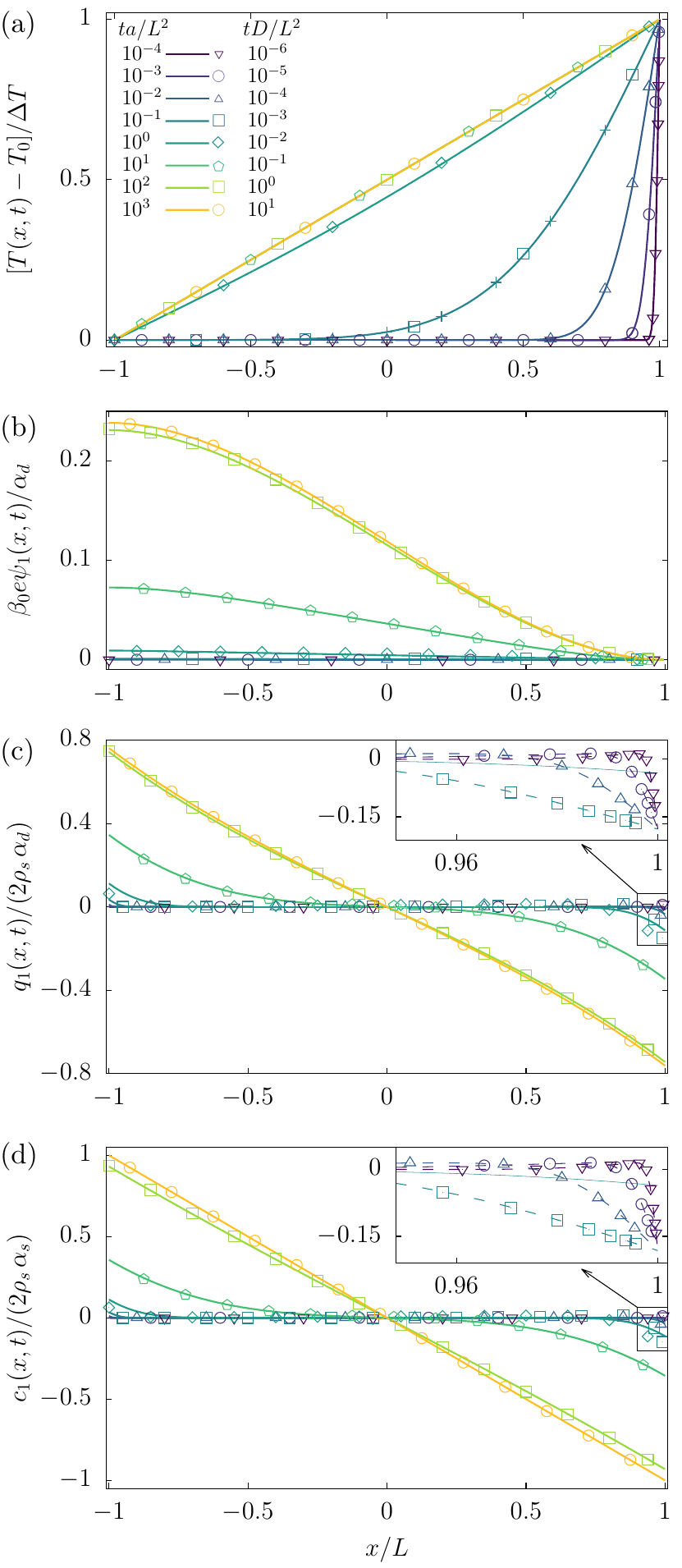}
         \caption{Thermal and ionic relaxation at $n=1$ and ${a}/D=100$ in response to a thermal quench
         ($\epsilon=10^{-3}$) at $x=L$.
         Numerical solution (symbols) to \eqr{eq:dimensionlessequations} at times $tD/L^2=\{10^{-6}- 10\}$ for $T$ (a), 
         $\psi$ (b), $q$ (c), and $c$ (d) where obtained at $\xi=1$, $\alpha_{+}=0.5$, $\alpha_{-}=0.1$, $f=2\times10^{-3}$.
         (a) Also shows \eqr{eq:pureTrelaxation} (lines) at the same times and \eqr{eq:pureTrelaxationearlyt} (plusses) at $tD/L^2=10^{-3}$.
         (b)-(d) Also show \eqr{eq:exactsolution} with lines. The insets of (c) and (d) show \eqr{eq:earlyq} (dashed lines) at times $tD/L^2=\{10^{-6}- 10^{-3}\}$}
    \label{fig:Tpsiqcn1response}
\end{figure}

\subsection{Local fields at $n=1$}
We show analytical (lines) and numerical (symbols) solutions to \eqr{eq:dimensionlessequations} for  
$n=1$ in Fig.~\ref{fig:Tpsiqcn1response}, where we plot the position dependence of $T$, $\psi$, $q$, and $c$ 
for logarithmically separated times between $tD/L^2=10^{-6}$ and $tD/L^2=10$. 
In Fig.~\ref{fig:Tpsiqcn1response} (a) we show the temperature. 
As a sanity check, we also compared numerical solutions for $(T(x,t)-T_{0})/\Delta T$ at $f=0$ to the exact 
result \eqr{eq:pureTrelaxation} [in this section we truncate the sums in Eqs.~\eqref{eq:pureTrelaxation} and \eqref{eq:exactsolution} after 2000 terms]: The difference 
between either predictions was at most $0.02$ (at $tD/L^2=10^{-6}$), dropping to $10^{-12}$ at late times.
The difference between $T$ calculated with either 
 $f=2\times10^{-3}$ or $f=0$ (and other parameters as before) was too small to detect within this numerical error margin. In any case, with our choice $f=2\epsilon$, the source term
 of the heat equation \eqref{eq:heatdimless} is $\mathcal{O}(\epsilon^3)$. Therefore, its effects are beyond the range of validity of our theory.
 
As anticipated in Sec.~\ref{Sec:purethermalrelaxation}, Fig.~\ref{fig:Tpsiqcn1response}(a) moreover shows that \eqr{eq:pureTrelaxationearlyt}  accurately describes 
 $T(x,t)$ at $ta/L^2=0.1$ (plusses) as well as at earlier times (not shown).
For the stated parameter set, Fig.~\ref{fig:Tpsiqcn1response} shows that $T(x,t)$ relaxes almost completely before 
$\psi, q$, and $c$ deviate from their initial values. Consequently, the ionic relaxation falls predominantly 
in the late-time regime ($t\gg\tau_T$) discussed in Sec.~\ref{sec:latetime}:
From $ta/L^2=10$ onwards, the assumption of a thermal steady state that we used to derive \eqr{eq:exactsolution} is justified.
Consequently, at late times, we observe a decent correspondence between numerics and the analytical predictions for $\psi_{1} ,q_{1}$, and $c_{1}$
[Eqs.~\eqref{eq:exactpotential}, \eqref{eq:exactchargedensity}, and \eqref{eq:exactconcentration}, respectively].  
Conversely, at early times ($t\ll\tau_{T}$), when \eqr{eq:pureTrelaxationearlyt}
accurately describes $T(x,t)$, one expects the predictions 
of \eqr{eq:earlyq} for $q_{1}$ and $c_{1}$ to be accurate.
Indeed, the inset of  Fig.~\ref{fig:Tpsiqcn1response}(c) (a zoom-in of the main panel to the region $x\lessapprox L$) 
shows an excellent agreement between \eqr{eq:earlyq} (dashed lines) and the same numerical data until $tD/L^2=10^{-3}$, while at that same time, \eqr{eq:exactchargedensity} 
gives erroneous predictions (the line does not pierce the open squares).
Interestingly, this inset exhibits
a tiny ionic charge density wave that moves with the front of thermal perturbation and
that breaks the antisymmetry (present at late times) of $q$ and $c$ around the midplane at early times. 
Given the equivalence at early times of $q_{1}/\alpha_{d}$ to
$c_{1}/\alpha_{s}$ as discussed in Sec.~\ref{sec:earlytime}, the inset of Fig.~\ref{fig:Tpsiqcn1response}(d) shows that the same analytical expression \eqr{eq:earlyq}
also describes the evolution of $c_{1}$ at early times well.

\subsection{Local fields at $n=100$}
\begin{figure}
\includegraphics[width=8.6cm]{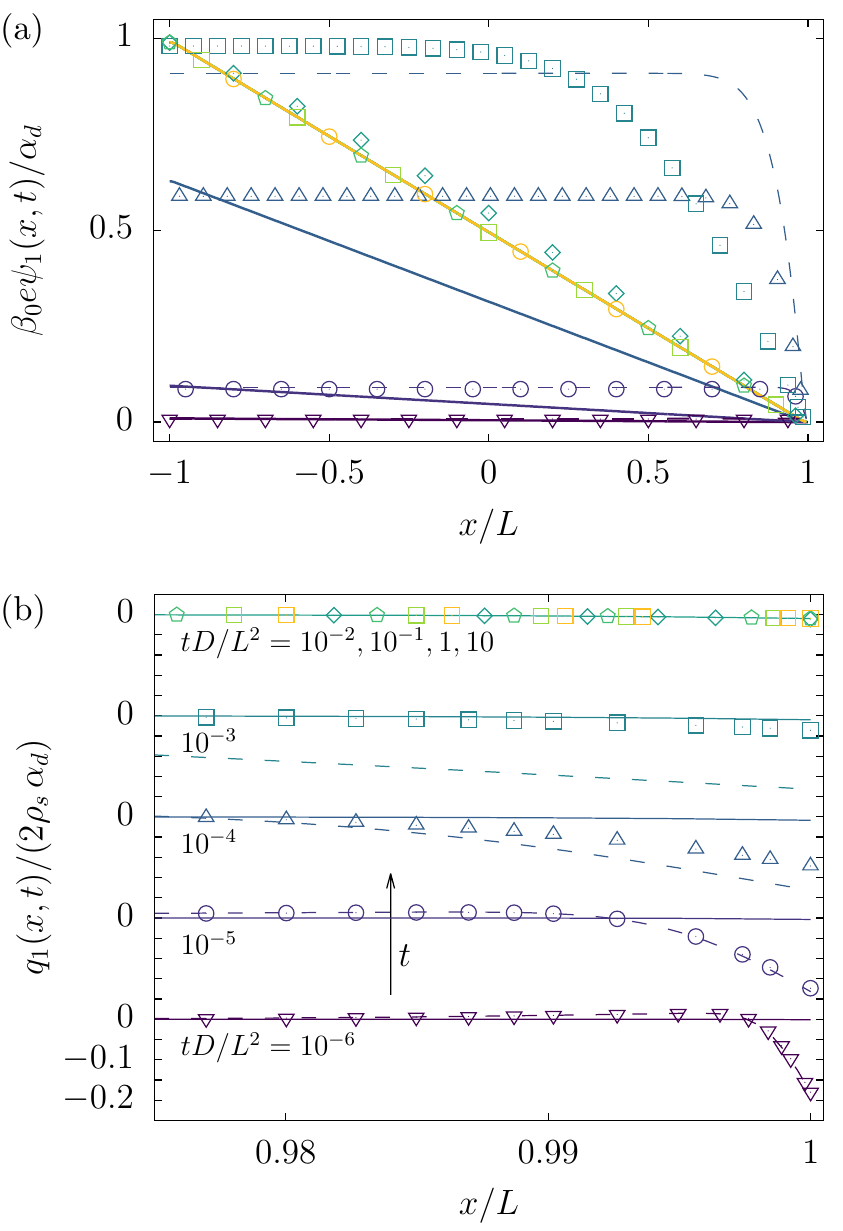}
         \caption{Numerical (symbols) and analytical (lines) results for  $\psi_{1}$  (a) and  $q_{1}(x,t)$ (b) at $n=100$. 
         All other parameters, colors, symbols, and line styles are as in Fig.~\ref{fig:Tpsiqcn1response}.}
      \label{fig:psiqcn100}
\end{figure}

Figure~\ref{fig:psiqcn100} shows numerical solutions to \eqr{eq:dimensionlessequations} 
and the same analytical approximations as before, now for $n=100$.
Since $T$ and $c_{1}$ are essentially $n$ independent [cf.~Eqs.~\eqref{eq:pureTrelaxation}, \eqref{eq:pureTrelaxationearlyt}, \eqref{eq:earlyq} and \eqref{eq:exactsolution}], we only show
 $\psi_{1}$ in Fig.~\ref{fig:psiqcn100}(a) and $q_{1}$ in Fig.~\ref{fig:psiqcn100}(b). 
We see in Fig.~\ref{fig:psiqcn100}(b) that \eqr{eq:earlyq} is now accurate only until $tD/L^2=10^{-5}$
[as this equation is $n$ independent, the dashed lines in Fig.~\ref{fig:psiqcn100}(b) are the same as in the inset of Fig.~\ref{fig:Tpsiqcn1response}(c)].
This difference with the $n=1$ case (accurate until $tD/L^2=10^{-3}$) is understood in terms of the larger error made for higher $n$
in neglecting the term $n^2\tilde{q}_{1}$ in \eqr{eq:NPearlyqt}.
Equation~\eqref{eq:exactchargedensity} is accurate after $tD/L^2=10^{-2}$, comparable to $tD/L^2=10^{-1}$ for the $n=1$ case.
 The key difference with the 
$n=1$ case, however, is that at $tD/L^2=10^{-1}$, $q_{1}(x,t)$ has already 
reached its steady-state profile [cf. \eqr{eq:ratiotimescales}: With increasing $n$, the early-time ($t\ll\tau_{T}$) regime of the transient response of $q$ and $\psi$ gains in importance].
Hence, \eqr{eq:exactchargedensity} is irrelevant
for the description of the \textit{transient} behavior of $q(x,t)$ for $n=100$ and solely captures its steady state. 
Yet, out of curiosity, we plot the corresponding late-time expression for $\psi_{1}$ [\eqr{eq:exactpotential}] in Fig.~\ref{fig:psiqcn100}(a);
while this expression gets the shape of $\psi_{1}$ completely 
wrong (except at steady state), it
 surprisingly accurately estimates the thermovoltage $V_{T}(t)=\epsilon\psi_{1}(-L,t)+\mathcal{O}(\epsilon^2)$ at all times considered.
Apparently, for the development of $V_{T}(t)$, it is not necessary that the thermal perturbation has spanned the system:
The local charge separation as observed in Fig.~\ref{fig:psiqcn100}(b) leads to the same voltage drop, 
but now already over the small region coincident with the thermal perturbation. 
Meanwhile, it comes as somewhat of a surprise that 
the analytical prediction for $\psi_{1}$ calculated  with Eqs.~\eqref{eq:earlyq}
provides fair approximations to our numerical results 
only for very early times ($tD/L^2=10^{-6}, 10^{-5}$), thereafter overestimating $\psi_{1}$ greatly. 
Since this method to approximate $\psi_{1}$ already goes awry at $t\ll\tau_{T}$, discrepancies cannot be attributed to 
usage of the approximate early-time temperature [\eqr{eq:pureTrelaxationearlyt} 
instead of \eqr{eq:pureTrelaxation}] in the derivation of \eqr{eq:earlyq}. 
Apparently, $\psi_{1}$ is very sensitive to the errors in $q_{1}$ (observable in Fig.~\ref{fig:psiqcn100}(b) from $tD/L^2=10^{-4}$ onwards) 
 resulting from the omission of the electromotive term $\tilde{c}_{0}\partial^{2}_{\tilde{x}}\tilde{\psi}_{1}$ in \eqr{eq:NPearlyt}.

\subsection{Boundary value relaxation}
\begin{figure}
    \centering
\includegraphics[width=8.6cm]{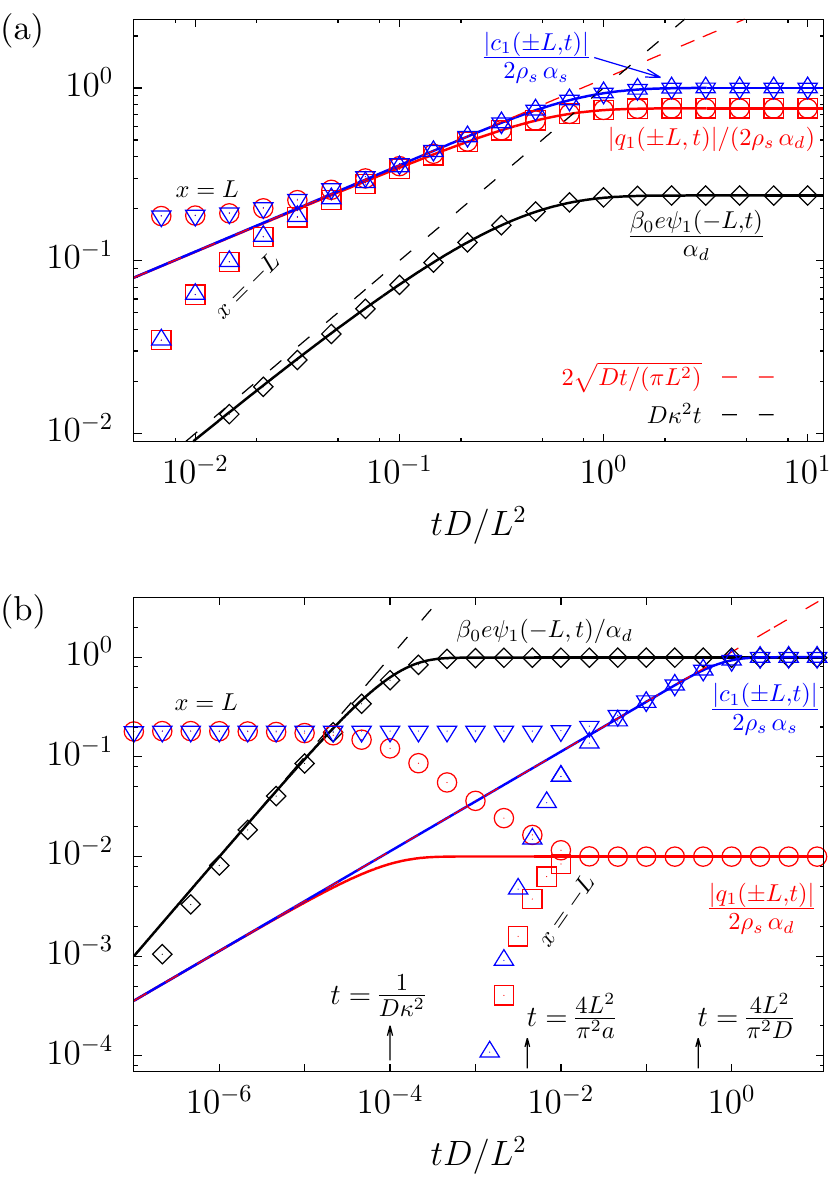}
       \caption{The relaxation of $\psi_{1}(-L,t), q_{1}(\pm L,t)$, and $c_{1}(\pm L,t)$ (black,  red, blue) from numerics 
       (symbols) and \eqr{eq:exactsolution} (lines) ($j\le10$), for $n=1$ (a) and $n=100$ (b).  
       Plotted as well are $D\kappa^2 t$ (black dashed) and $2\sqrt{Dt/(L^2\pi)}$ (red dashed).}
      \label{fig:boundaryresponse}
\end{figure}

In Fig.~\ref{fig:boundaryresponse} we show numerics (symbols) and analytical predictions from \eqr{eq:exactsolution} (lines)
for the relaxation of $\psi_{1}(-L,t)$ 
and the absolute boundary values of the ionic charge and salt densities, $|q_{1}(\pm L,t)|$ and $|c_{1}=\pm L,t)|$, respectively. 
Concerning $V_{T}(t)=\epsilon\psi_{1}(-L,t)+\mathcal{O}(\epsilon^2)$, we see that analytical predictions agree well with numerics 
for $n=1$ [unsurprising, given the agreement observed in Fig.~\ref{fig:Tpsiqcn1response}(b)] and $n=100$, where a minor 
discrepancy is observed at very early times.
Again, since the steady-state temperature ansatz is only justifiable after $tD/L^2=0.1$, 
the good agreement observed Fig.~\ref{fig:boundaryresponse}(b) between numerics and \eqr{eq:exactpotential} up to $tD/L^2=10^{-6}$ 
[pushing that equation five orders of magnitude into temporal terra incognita] is
remarkable.
The small-$t$ scaling of $\beta_{0} e V_{T}/(\alpha_{d}\epsilon)=\kappa^2 Dt$ (black dashed lines) is derived in Appendix~\ref{Appendix:smallt}. 

We see in Fig.~\ref{fig:boundaryresponse} that, at early times ($t<\tau_{T}$),
$|c_{1}(\pm L,t)|$ and $|q_{1}(\pm L,t)|$ are perturbed at the quenched ($x=L$) electrode (down triangles and circles), 
and unperturbed at the other side (up triangles and squares).
With \eqr{eq:earlyq} we find that the early-time plateaus observed in Fig.~\ref{fig:boundaryresponse} lie at 
$\lim_{t\to0^{+}}|q_{1}(L,t)|/(2\rhos \alpha_{d})=\lim_{t\to0^{+}}|c_{1}(L,t)|/(2\rhos \alpha_{s})=2/(1+\sqrt{a/D})\approx 0.18$, independent of $n$. 
Interestingly, at $n=100$ this prediction for $c_{1}$ is still accurate around $t\approx\tau_{T}$, when \eqr{eq:earlyq} inaccurately
describes $q_{1}$ [Fig.~\ref{fig:psiqcn100}(b)]. At $n=1$ the omission of the $n^2 q_{1}$ term in \eqr{eq:NPearlyqt} is justifiable and the prediction
$\lim_{t\to0^{+}}|q_{1}(L,t)|/(2\rhos \alpha_{d})\approx 0.18$ holds up to $t\approx\tau_{T}$ as well.
From $tD/L^2=0.01\Leftrightarrow t\approx\tau_{T}$ onwards, we see, for $n=1$, that numerics and analytical predictions 
from \eqr{eq:exactsolution} converge, in line with our observations in Fig.~\ref{fig:Tpsiqcn1response}.
Once converged, they scale as  $|q_{1}(\pm L)|= |c_{1}(\pm L)|=2\sqrt{Dt/(L^2\pi)}$ (red dashed lines) as derived in Appendix~\ref{Appendix:smallt}, 
finally relaxing to their steady-state values around $tD/L^2=1\Leftrightarrow t\approx\tau_{q}$.

For $n=100$ and $a/D=100$, \eqr{eq:exactsolution} predicts that $V_{T}(t)$ and $q_{1}(\pm L,t)$ relax  
 two orders of magnitude faster [at $t=L^2/(D[n^2+\pi^2/4])\approx 1/(D\kappa^2)$] than $T$ [at $t=4L^2/(a \pi^2)$].
While $V_{T}(t)$ really does develop on this short timescale (as discussed above),  $q_{1}(\pm L,t)$ becomes enslaved to the ``slow" thermal relaxation. 
Together with relaxation of $c_{1}(\pm L,t)$ at $t= 4L^2/(D \pi^2 )$, Fig.~\ref{fig:boundaryresponse}(b) 
shows a separation of timescales over four orders of magnitude for the 
three observables $\psi_{1}(-L,t), |q_{1}(\pm L,t)|$, and $|c_{1}(\pm L,t)|$.
The separation of timescales of boundary observables $V_{T}(t)$ and $q(\pm L,t)$ seems to contradict the intuition 
that ionic charge and electrostatic potential are instantaneously related via
the Poisson equation, and should thus relax in lockstep. 
However, the Poisson equation is a \textit{nonlocal} relation between $\psi(x,t)$ and $q(x,t)$, which, apparently,
does not forbid $\psi$ and $q$ to relax differently at specific locations.
Indeed, Fig.~\ref{fig:psiqcn100}(a) clearly shows that the overall electrostatic potential 
$\psi_{1}(x,t)$ reaches its steady state much later (around $ta/L^2=1$) than $\psi_{1}(-L,t)$.

\section{Unequal ionic diffusivities}\label{sec:unequalD}
We note that our finding $V_{T}(t)\sim \exp[-D\kappa^2t]$ in
Fig.~\ref{fig:boundaryresponse}(b) is at odds with the experimental data of 
\cit{bonetti2015thermoelectric}. They studied a $6$-mm-wide cell filled with a concentrated electrolyte 
(2M EMIMTFSI in acetonitrile) subject to a thermal quench of $\Delta T=20$ K. 
Their measurements indicated that $V_{T}(t)\sim \exp[-\pi^{2}D_{+}t/(4L^2)]$, where the fitted cationic diffusion constant
$D_{+}$ was a factor 3 off from
literature values for the pure EMIMTFSI ionic liquid (without solvent). 
In principle, the discrepancy between our works could have arisen due to several 
simplifying assumptions underlying our model, as their setup: 
(1) used a concentrated electrolyte, for which our continuum Nernst-Planck 
description of the ionic currents (reasonable for dilute electrolytes) 
might be unsuitable and to which it is difficult to assign a Debye length; 
(2) had comparable lateral and in-plane dimensions, which further undermines our one-dimensional model; and
(3) was exposed to a thermal quench two orders of magnitude larger than what we imposed in Sec.~\ref{sec:Results}.
Accordingly, we performed exploratory numerical simulations of \eqr{eq:dimensionlessequations} 
with $\epsilon=0.1$ (and $T$ independent dimensionless parameters) and found that the third speculation 
does not explain the discrepancy: In that case, the qualitative behavior 
[including the fast $V_{T}(t)\sim \exp[-D\kappa^2t]$-relaxation]
of our model is unaltered, with the notable exception that the
antisymmetry of the steady-state profiles of $c$ and $q$ around $x=0$ is broken.

Instead of the above three speculations, it turns out that the qualitative features of
the experimental data of \cit{bonetti2015thermoelectric} \textit{can} be
reproduced by our model if one accounts for different diffusivities among the ions \cite{private}. After all, NMR measurements of pure EMIMTFSI (without solvent) 
determined an apparent cationic transference number $D_{+}/(D_{-}+D_{+})\approx 0.63$, which implies $\xi=1.7$ \cite{noda2001pulsed}.

Once more using the steady-state temperature ansatz, in Appendix~\ref{AppendixC} we derive $V_{T}$ for $n\gg1$ and for general $\xi$:
\begin{widetext}
\begin{align}\label{eq:VTfinal}
\frac{eV_{T}(t)}{\kbt_{0} \epsilon}&=\alpha_{\rm d}-2\alpha_{\rm s}\frac{1-\xi}{1+\xi}\sum_{j\ge1}\frac{1}{\mathcal{N}_{j}^2}\exp{\left[-\frac{2\mathcal{N}_{j}^2}{1+\xi}\frac{D_{+}t}{L^2}\right]}%\nn
%&\quad
-4\frac{\xi\alpha_{+}-\alpha_{-}}{1+\xi}\sum_{j\ge1}\frac{1}{\mathcal{N}_{j}^2}\exp{\left[-\frac{1+\xi}{2\xi}D_{+}\kappa^2 t\right]}%\nn
%&\quad
+\mathcal{O}\left(n^{-1}\right)+\mathcal{O}\left(\epsilon\right)\,,
\end{align}
\end{widetext}
which reduces correctly to the $n\gg1$ limit of $\psi_{1}(-L)$ [cf. \eqr{eq:thermovoltagefastthermal}] for $\xi=1$. 
Strikingly, in \eqr{eq:VTfinal} now appear relaxation times that scale like the diffusion time as $\sim L^2$, which is a promising sign 
for our attempt at explaining the data of \cit{bonetti2015thermoelectric}. Moreover, we can rewrite the exponents of \eqr{eq:VTfinal} to $\exp{\left[-D_{a}\kappa^2 t  \right]}$ and 
$\exp{\left[-\mathcal{N}_{j}^2 D_{h} t/L^2  \right]}$, respectively, with $D_{a} \equiv (D_{-}+D_{+})/2$ being the arithmetic and $D_{h} \equiv 2/[(1/D_{-}+1/D_{+})]$ being 
the harmonic mean of the ionic diffusion constants, respectively \cite{alexe2007relaxation, balu2018role}. Notably, precisely these two means appear in the electrolyte conductance and in Nernst's expression for the ambipolar diffusivity of neutral salt, respectively. 
While it is now tempting to interpret $V_{T}(t)$ as being generated simultaneously by ionic charge density \textit{and} salt density relaxation, we note that $V_{T}(t)$ is ultimately only directly related to $q(x,t)$ (cf. Appendix~\ref{AppendixC}). 
The appearance of parameters typical for salt diffusion ($\alpha_{s}$ and $L^2/D_{h}$) merely suggests that there is
a nontrivial coupling between $c$ and $q$ whenever $\xi\neq1$. 
We leave an in-depth analysis of $q$ and $c$ at $\xi\neq 1$ for future work.

\begin{figure}
\includegraphics[width=8.6cm]{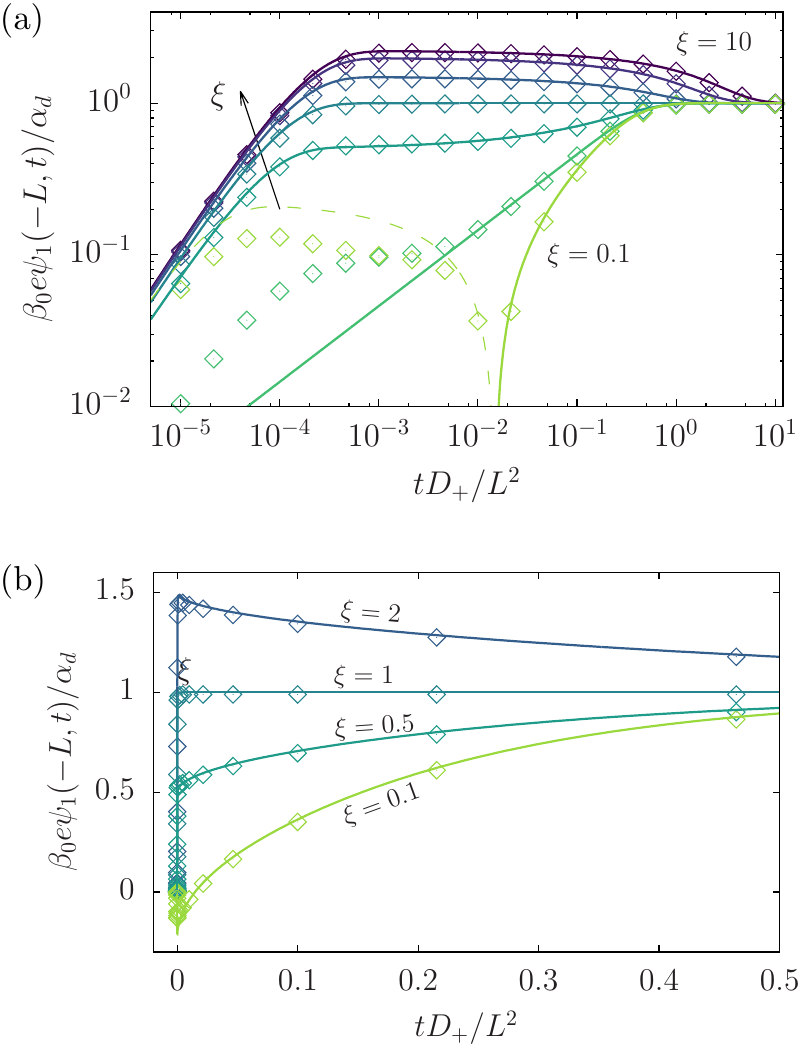}
      	\caption{Predictions for $V_T(t)=\epsilon \psi_{1}(-L,t)+\mathcal{O}(\epsilon^2)$ from a numerical simulation of \eqr{eq:dimensionlessequations} (symbols)  and from the analytical expression \eqr{eq:VTfinal} (lines), on double logarithmic scales (a) and linear scales (b). The dashed line represents $- \psi_{1}(-L,t)$ at times where $ \psi_{1}(-L,t)<0$.  In (a), 
      	going from top to bottom, the lines represent $\xi=\{10, 5, 2, 1, 0.5, 0.2, 0.1\}$. All other parameters are the same as in Fig.~\ref{fig:Tpsiqcn1response}.      	}
    \label{fig:VT}
\end{figure}

In Fig.~\ref{fig:VT}, we plot \eqr{eq:VTfinal} (lines) for $n=100$ and several $\xi=\{10, 5, 2, 1, 0.5, 0.2, 0.1\}$, and all other parameters 
the same as in Sec.~\ref{sec:Results}.
Overall, we observe a good agreement between that equation 
and numerical simulations of  \eqr{eq:dimensionlessequations} (symbols).
Noticeable deviations occur for small $\xi$ and $t<\tau_{T}$, when the steady-state temperature ansatz used 
to derive \eqr{eq:VTfinal} is unjustified.
Note, also,  that the $\xi=1$-case corresponds to the black diamonds in Fig.~\ref{fig:boundaryresponse}(b). 
For $\xi\neq1$, the data plotted on double logarithmic scales [Fig.~\ref{fig:VT}(a)] exhibits two distinct relaxation processes: 
a quick rise of $V_{T}$ on the Debye timescale, followed by a slower $L^2/D_{+}$ relaxation towards
the steady state.
In between these two timescales, $V_{T}$ exhibits a plateau, whose height $V^{p}_{T}$ can be found by setting $tD_{+}/L^2=0$ and $tD_{+}\kappa^2=\infty$ in \eqr{eq:VTfinal}:
\begin{align}\label{L2DVT}
\frac{eV_{T}^{p}}{\kbt_{0} \epsilon}=\frac{2}{1+\xi}\left[\xi\alpha_{+}-\alpha_{-}\right]\,.
\end{align}
In fact, for $tD_{+}\kappa^2=0$, both sums in \eqr{eq:VTfinal} can be performed 
and \eqr{eq:VTfinal} correctly predicts $V_{T}(0)=0$. 
For the case of KCl as discussed in Sec.~\ref{sec:Results}, Eqs.~\eqref{eq:VTfinal} and \eqref{L2DVT} indicate that in Fig.~\ref{fig:boundaryresponse}(b) we missed 
an intermediate-time voltage plateau $2\%$ below  the steady-state thermovoltage, and the slow $L^2/D_{+}$ relaxation from one to the other.
        
When plotted along linear axes [Fig.~\ref{fig:VT}(b)], 
$V_{T}$ seemingly instantaneously jumps to the aforementioned plateau values, and relaxes to the steady state thereafter.
Here, the case $\xi=0.5$ looks
similar to the data of Figs.~2 and 3 of \cit{bonetti2015thermoelectric}. A quantitative
comparison between our works is not possible, however, as there is no data for $\xi$ and $\alpha_{\pm}$ 
of EMIMTFSI in acetonitrile at the dilution used in \cit{bonetti2015thermoelectric}. 
The fact that $V_{T}$ ``overshoots'' its steady-state value for certain combinations of $\xi$'s and $\alpha_{\pm}$'s could be exploited  
 to boost the performance of thermally chargeable capacitors.    
With \eqr{L2DVT} and the tabulated data of \cit{agar1989single}
we see that alkali hydroxides and hydrohalic acids could be promising electrolytes for this purpose 
[for example, $eV_{T}^{p}/(\kbt_{0} \epsilon)=-5.75$ for LiOH].
Alternatively, with knowledge of the steady-state thermovoltage, the intermediate-time thermovoltage plateau value, and 
either $\xi$, $\alpha_{+}$ or $\alpha_{-}$, one can give an indirect prediction of the other two.

\section{Discussion}\label{sec:Discussion}
Recent molecular dynamics simulation of a binary mixture subject to a thermal quench 
have predicted an early-time local mole fraction (Fig.~5 in \cit{hafskjold2017non}) 
very similar to the density profile in Fig.~\ref{fig:psiqcn100}(b).
For these uncharged molecules, the absence of an electromigration term 
in the fluxes is obvious \cite{de1942theorie, bierlein1955phenomenological,costeseque2004transient}. 
Hence, it would be interesting to see to what extend \eqr{eq:earlyq} describes the early-time 
thermodiffusion of binary mixtures as well. 

Moreover, \eqr{eq:earlyq} sheds new light on an age-old puzzle, the very fast temperature-induced concentration polarization observed
by Tanner in 1927 \cite{tanner1927soret}. 
Our analytical result
$\lim_{t\to0^{+}}|c_{1}(L,t)|/(2\rhos\alpha_{s})=2/(1+\sqrt{a/D})$ and numerical data in Fig.~\ref{fig:boundaryresponse} naturally
indicate that a nonzero boundary salt density is present for all nonzero times.
These results complement earlier efforts \cite{thomaes1951recherches,van1990kinetics} to explain Tanner's observations with calculations
that used the steady-state temperature profile $(T-T_{0})/\Delta T =1/2+x/2L$ at all times. 

\section{Conclusion}\label{sec:Conclusion}
We have studied the response of a  model electrolytic cell subject to a quench in the temperature 
at one of its two confining electrode surfaces. 
The system is modeled by four coupled differential equations [\eqr{eq:dimensionlessequations}] 
and boundary conditions [\eqr{eq:dimlessinitandcond}] in which 
seven dimensionless numbers appear: the size of the quench $\epsilon$, the Debye 
separation parameter $\kappa L$,  the ratio of ionic diffusivities $D_{+}/D_{-}$,
the ratio of thermal to cationic diffusivities 
${a}/D_{+}$, the reduced ionic Soret coefficients $\alpha_{+}$ and $\alpha_{-}$, and the combination $\kb\rhos/\varrho c_{p}$ for ionic heat production, respectively. 

We first studied the case $D_{+}=D_{-}$, which is relevant to, e.g., aqueous KCl, RbBr, RbI, and CsBr. In this case we found analytical approximations to the ionic charge density $q$, neutral salt concentration $c$,
and electrostatic potential $\psi$ for early and late times compared to the thermal relaxation.
These expressions were shown to correspond well to numerical simulations of the same quantities in their 
respective temporal regimes of validity [we performed the numerical simulations of \eqr{eq:dimensionlessequations} using a parameter set typical for aqueous KCl]. 
This leaves behind an intermediate time window for which we only have numerical data. 
Notably, the size of this window depends on $\kappa L$ because the early-time expression for $q$ was
derived with the omission of the thermodiffusion term $(\kappa L)^2 q$ in the ionic charge current. 
This means that the early-time expressions 
approximate $q$ over a longer time period at $\kappa L=1$ (valid until $tD/L^2=10^{-3}$) 
than at $\kappa L=100$ (valid until $tD/L^2=10^{-5}$).
The importance of either regimes (early- and late-time) was shown to depend on $\kappa L$. 
For $\kappa L=1$, the system behaves mainly as explained in  Ref.~\cite{stout2017diffuse}:
The quenched temperature relaxes quickly, after which the electrostatic potential and ionic charge and salt densities relax slowly.
Conversely, for $\kappa L=100$, the rearrangement of ions in thermal gradients is
sufficiently fast that the ionic charge density can track the thermal relaxation. 
For all parameters considered, an ionic charge density wave is observed that spreads 
as the thermal perturbation travels through the system.
While the ionic relaxation becomes enslaved to the slow thermal relaxation, the thermovoltage develops 
on the Debye timescale,  the fastest timescale of the system.

For the case of $D_{+}\neq D_{-}$, we have shown that the relaxation of the thermovoltage happens via a two-step process: a fast relaxation on the Debye timescale,
followed by a slower diffusive relaxation. In fact, for a suitably chosen electrolyte, the thermovoltage overshoots 
its steady-state value. This feature could be exploited for enhanced thermal energy scavenging by thermally chargeable capacitors.

The main conclusions of this article are twofold: Depending on the Debye separation parameter, (1) 
 assuming an instantaneous steady-state temperature profile leads to satisfactory predictions for the transient salt density profiles, 
but wrong predictions for the transient ionic charge density and electrostatic potential profiles;
 (2) the thermovoltage relaxes both on the Debye timescale $1/(D\kappa^2)$ and the
  diffusion timescale $L^2/D$. The relative importance of these two relaxation 
  processes depends on the ionic Soret coefficients and on the ratio of ionic diffusivities.

\begin{acknowledgments}
MJ thanks Joost de Graaf, Sviatoslav Kondrat, Paolo Malgaretti,
Marco Bonetti, Sawako Nakamae, and Michel Roger for stimulating discussions and for useful comments on our manuscript.
\end{acknowledgments}

\begin{appendix}

\section{Derivation of \eqr{eq:exactsolution}}\label{sec:exactinversion}
We apply 
Laplace transformations on Eqs.~\eqref{eq:Poissondimless} and \eqref{eq:electrokineticslowions} to transform
the PDEs for $\tilde{\psi}_{1}$, $\tilde{q}_{1}$, and $\tilde{c}_{1}$ into ODEs
for their Laplace transformed counterparts $\hat{\tilde{\psi}}_{1}$, $\hat{\tilde{q}}_{1}$,
and $\hat{\tilde{c}}_{1}$ [we denote the Laplace transform of a function $f(\tilde{x},\tilde{t}\,)$ 
by $\hat{f}(\tilde{x},s)=\int_{0}^{\infty}\textrm{d}\tilde{t}\,\exp{(-s\tilde{t}\,)}f(\tilde{x},\tilde{t}\,)$].
Stout and Khair \cite{stout2017diffuse} already found solutions to these ODEs [see their Eq.~(20)], which in our notation read
\begin{subequations}\label{eq:laplaceperturb}
\begin{align}
\frac{\hat{\tilde{c}}_{1}(\tilde{x},s)}{\alpha_{\rm s}}&=-\frac{1}{rs}\frac{\sinh (r\tilde{x})}{\cosh r}\,,\label{eq:laplacec}\\
\frac{\hat{\tilde{q}}_{1}(\tilde{x},s)}{\alpha_{\rm d}}&=-\frac{1}{ms}\frac{\sinh (m\tilde{x})}{\cosh m}\,,\label{eq:laplaceq}\\
\frac{\hat{\tilde{\psi}}_{1}(\tilde{x},s)}{\alpha_{\rm d}}&=\frac{n^2}{2m^2 s}\left[\frac{\sinh (m\tilde{x})-\sinh m}{m\cosh m}+1-\tilde{x}\right]\label{eq:laplacef},
\end{align}
\end{subequations}
with $r^2=s$ and $m^2=n^2+s$. 
To determine $\tilde{\psi}_{1}, \tilde{q}_{1}$, and $\tilde{c}_{1}$, we need to perform inverse Laplace transformations on \eqr{eq:laplaceperturb}.
For instance, determining $\tilde{\psi}(\tilde{x},\tilde{t}\,)$ comes down to
\begin{align}\label{eq:ftot}
\hat{\tilde{\psi}}_{1}(\tilde{x},\tilde{t}\,)&=\sum_{s\in s_{\ell}}\text{Res}\left(\hat{\tilde{\psi}}_{1}\exp(s\tilde{t}\,),s_{\ell}\right)\,,
\end{align}
where the poles $s_{\ell}=\{s_{0}, s_{n}, s^{\star}_{j} \}$ of $\hat{\tilde{\psi}}_{1}(\tilde{x}, s)$ are located at
$s_{0}=0$, $s_{n}=-n^{2}$, and $s^{\star}_{j}=(m^{\star}_{j})^2-n^2$ where 
$m^{\star}_{j}=\pm i (j-1/2)\pi\equiv\pm i\mathcal{N}_{j}$ with $j\in\mathbb{N}$.

The pole $s_{0}=0$ gives the steady-state solution,
\begin{align}\label{eq:f0}
\text{Res}\left(\hat{\tilde{\psi}}_{1}\exp(s\tilde{t}\,),0\right)=\frac{\sinh (n \tilde{x})-\sinh n}{2n\cosh  n }+\frac{1-\tilde{x}}{2}.
\end{align}
To find the residue of the pole at $s_{n}=-n^{2}$, we expand $\hat{\tilde{\psi}}_{1}$ around $s=-n^{2}$,
\begin{align}\label{eq:fkd}
\hat{\tilde{\psi}}_{1}\overset{s\to-n^2}=&\frac{1}{2m^2 }\left[\cancel{\frac{x}{L}-\frac{x}{L}} + 
\mathcal{O}\left(m^3 \right)\right]=\mathcal{O}\left(m \right)\,.
\end{align}
This implies
\begin{align}\label{eq:fkd2}
\text{Res}\left(\hat{\tilde{\psi}}_{1}\exp(s\tilde{t}\,),s=-n^{2}\right)=0,
\end{align}
because $m=0$ at $s=-n^2$.
For the poles at $s^{\star}_{j}$ we expand 
\begin{align}\label{eq:A6}
\cosh (m)\overset{s\to s^{\star}_{j}}=& \frac{\sinh m}{2m}\bigg|_{s=s^{\star}_{j}}\left(s- s^{\star}_{j}\right)\nn
\Rightarrow \frac{1}{\cosh (m)}\overset{s\to s^{\star}_{j}}=& \frac{2i(-1)^{j}m^{\star}}{s-s^{\star}_{j}}\,,
\end{align}
where, going to the second line we used $m(s^{\star}_{j})=\pm m^{\star}_{j}$, and $\sinh m^{\star}_{j}=i(-1)^{j+1}$. 
We find
\begin{align}\label{eq:fstar}
&\sum_{j\ge1}\text{Res}\left(\hat{\tilde{\psi}}_{1}\exp(s\tilde{t}\,),s^{\star}_{j}\right)=\nn
&=\sum_{j\ge1}
\text{Res}\left(\frac{i(-1)^{j}[\sinh (m^{\star}\tilde{x})-\sinh m^{\star} ] }{(m^{\star}/n)^{2} s^{\star}}\frac{\exp(s\tilde{t}\,)}{s- s^{\star}},s^{\star}_{j}\right)\nn
 &=-\sum_{j\ge1}\frac{1+(-1)^{j}\sin(\mathcal{N}_{j}\tilde{x})}
{\mathcal{N}_{j}^{2}\left[1+\mathcal{N}_{j}^{2}/n^2\right]}\exp{\left[-\left( n^{2}+\mathcal{N}_{j}^{2}\right)\tilde{t}\,\right]}.
 \end{align}
Combining Eqs.~\eqref{eq:f0}, \eqref{eq:fkd}, and \eqref{eq:fstar} yields \eqr{eq:exactpotential}.

We now easily find $\tilde{q}_{1}$ [\eqr{eq:exactchargedensity}]  by inserting \eqr{eq:exactpotential} into the Poisson equation \eqref{eq:Poisson}.
Likewise, noting that  
the equations governing $\tilde{q}_{1}/\alpha_{\rm d}$ and $ \tilde{c}_{1}/\alpha_{\rm s}$
are the same for $n\to0$ [cf.~\eqr{eq:electrokineticslowions}],  $\tilde{c}_{1}$ [\eqr{eq:exactconcentration}] is found from  \eqr{eq:exactchargedensity} by taking $n\to0$ therein.
We have checked \eqr{eq:exactsolution} against numerical
Laplace inversions of Eq.~\eqref{eq:laplaceperturb}, 
using the 't Hoog algorithm \cite{de1982improved}. The results coincided perfectly for all times and parameters considered.

Before performing the inverse Laplace transforms on \eqr{eq:laplaceperturb}, \cit{stout2017diffuse} 
   first applied Pad\'{e} approximations to those expressions.
   Approximations to $\tilde{\psi}_{1}, \tilde{q}_{1}$, and $\tilde{c}_{1}$ are then easily read off.
Notably, the timescales $\tau^{\rm app}_{q}$ and $\tau^{\rm app}_{\psi}$  with which the approximated 
$\tilde{q}_{1}$ and $\tilde{\psi}_{1}$ relaxed were unequal, $\tau^{\rm app}_{q}\neq\tau^{\rm app}_{\psi}$. 
However, since $\hat{\tilde{q}}_{1}(\tilde{x},s)$ and $\hat{\tilde{\psi}}_{1}(\tilde{x},s)$ have the same pole structure,
any difference between $\tau^{\rm app}_{q}$ and $\tau^{\rm app}_{\psi}$ must stem from the 
Pad\'{e} approximation scheme employed. 
Other than fixing this glitch, the merits of \eqr{eq:exactsolution} over the approximate expressions of \cit{stout2017diffuse}
are limited: As discussed in \cit{janssen2018transient}, Pad\'{e} approximations around $s_{0}=0$ lead to decent predictions 
for the late-time response of the respective functions. Indeed, we have seen that \eqr{eq:exactsolution} 
(that also captures all fast-decaying $s_{j}^{\star}$ modes) deviates strongly from the Pad\'{e} approximations
only at early times ($\tilde{t}<0.1$). But as discussed in the main text, at those early times, \eqr{eq:exactsolution} does not describe the physics of interest, 
because the steady-state temperature ansatz is erroneous there.

\section{Early-time boundary value scaling of \eqr{eq:exactsolution}}\label{Appendix:smallt}
From \eqr{eq:exactpotential} follows a prediction for $V_{T}(t)$:
\begin{align}\label{eq:thermovoltagefastthermal}
\frac{\beta_{0}eV_{T}(\tilde{t}\,)}{\alpha_{\rm d}\epsilon}
&=1-\frac{\tanh n }{n}-2\sum_{j\ge1}\frac{\exp{\left[-\left( n ^{2}+\mathcal{N}_{j}^2\right)\tilde{t}\,\right]}}
{\mathcal{N}_{j}^2\left[ 1+\mathcal{N}_{j}^2/n ^2\right]}\nn
&\quad+\mathcal{O}\left(\epsilon\right)\,.
\end{align}
Expanding this expression around $\tilde{t}=0$,
for the first two terms of the expansion, the infinite sum can be performed. 
Thence, at short times, $V_{T}$ increases as
\begin{align}\label{eq:Vtsmallt}
\lim_{\tilde{t}\to0}\frac{\beta_{0}eV_{T}(\tilde{t}\,)}{\alpha_{\rm d}\epsilon}
&=n^2 \tilde{t}\,.
\end{align}

To determine $\lim_{\tilde{t}\to0} c(-L,\tilde{t}\,)$,
we rewrite \eqr{eq:exactconcentration} to 
\begin{align}\label{appeq:qsmallt}
\frac{\tilde{c}_{1}(-1,\tilde{t}\,)}{\alpha_{\rm s}}
&=2\sqrt{\tilde{t}}\sum_{j\ge1}\sqrt{\tilde{t}}\,\frac{1-\exp{\left[-p^2\right]}}{p^2}.
\end{align}
with $p=\mathcal{N}_{j} \sqrt{\tilde{t}}$.
Now consider the following integral
\begin{align}\label{eq:appBeq4}
 &\int_{\mathcal{N}_{j} \sqrt{\tilde{t}}}^{\mathcal{N}_{j+1} \sqrt{\tilde{t}}}\upd p\,\frac{1-\exp{\left[-p^2\right]}}{p^2}=\nn
 &\quad\quad=-\int_{\mathcal{N}_{j} \sqrt{\tilde{t}}}^{\mathcal{N}_{j+1} \sqrt{\tilde{t}}}\upd p\,\left(p-\mathcal{N}_{j+1} \sqrt{\tilde{t}}\right)\frac{\upd}{\upd p}\frac{1-\exp{\left[-p^2\right]}}{p^2}\nn
 &\quad\quad\quad-(\mathcal{N}_{j} -\mathcal{N}_{j+1} ) \sqrt{\tilde{t}}\,\frac{1-\exp{\left[-\mathcal{N}^2_{j} \tilde{t}\,\right]}}{\mathcal{N}^2_{j} \tilde{t}}\,,
 \end{align}
where we used integration by parts and $(\upd/\upd p)(p-j\pi\sqrt{\tilde{t}})=1$.
 With $\mathcal{N}_{j} -\mathcal{N}_{j+1}=-\pi$ we conclude  that
\begin{align}
\frac{1-\exp{\left[-\mathcal{N}^2_{j} \tilde{t}\,\right]}}{\mathcal{N}^2_{j}  \sqrt{\tilde{t}}}&=
\int_{\mathcal{N}_{j}\sqrt{\tilde{t}}}^{\mathcal{N}_{j+1}\sqrt{\tilde{t}}}\upd p\,\frac{1-\exp{\left[-p^2\right]}}{\pi p^2}
+\mathcal{O}(\tilde{t}\,)\,,
 \end{align}
because the integral on the right hand side of \eqr{eq:appBeq4} is $\mathcal{O}(\tilde{t}\,)$. 
Inserting the above result into \eqr{appeq:qsmallt} gives 
\begin{align}
\frac{\tilde{c}_{1}(-1,\tilde{t}\,)}{\alpha_{\rm s}}&=
\frac{2}{\pi}\sqrt{\tilde{t}}\sum_{j\ge1}\int_{\mathcal{N}_{j}\sqrt{\tilde{t}}}^{\mathcal{N}_{j+1}\sqrt{\tilde{t}}}\upd p\,\frac{1-\exp{\left[-p^2\right]}}{p^2}+\mathcal{O}(\tilde{t}\,^{3/2})\nn
&=
\frac{2}{\pi}\sqrt{\tilde{t}}\int_{\pi\sqrt{\tilde{t}}/2}^{\infty}\upd p\,\frac{1-\exp{\left[-p^2\right]}}{p^2}+\mathcal{O}(\tilde{t}\,^{3/2})\nn
&=
\frac{2}{\pi}\sqrt{\tilde{t}}\int_{0}^{\infty}\upd p\,\frac{1-\exp{\left[-p^2\right]}}{p^2}+\mathcal{O}(\tilde{t}\,)\nn
&=2\sqrt{\frac{\tilde{t}}{\pi}}+\mathcal{O}(\tilde{t}\,)\,.
\end{align}
With a similar calculation one finds  $\lim_{\tilde{t}\to0}\tilde{q}_{1}(-1,\tilde{t}\,)/\alpha_{\rm d}=2\sqrt{\tilde{t}/\pi}+\mathcal{O}(\tilde{t}\,)$.

\section{Derivation of \eqr{eq:VTfinal}}\label{AppendixC}
Similar to \eqr{eq:electrokineticslowions}, but now for $\xi\neq1$, we find 
\begin{subequations}\label{eq:3}
\begin{align}
\partial_{\tilde{t}} \tilde{\rho}_{+,1}&=\partial^2_{\tilde{x}} \tilde{\rho}_{+,1}-\frac{n^2}{2}\left( \tilde{\rho}_{+,1}- \tilde{\rho}_{-,1}\right) \,,\\
\xi\partial_{\tilde{t}} \tilde{\rho}_{-,1}&=\partial^2_{\tilde{x}} \tilde{\rho}_{-,1}+\frac{n^2}{2}\left( \tilde{\rho}_{+,1}- \tilde{\rho}_{-,1}\right)\,.
\end{align}
\end{subequations}
We apply Laplace transformations of both sides of \eqr{eq:3} and group the result in a matrix equation,
\begin{subequations}
\begin{align}
 \left( \begin{array}{c}  \partial^2_{\tilde{x}}\hat{\tilde{\rho}}_{+,1}\\ \partial^2_{\tilde{x}}\hat{\tilde{\rho}}_{-,1}  \end{array} \right) 
 =& \def\arraystretch{2.1}\begin{pmatrix} \displaystyle{s+\dfrac{n^2}{2}}&\displaystyle{-\dfrac{n^2}{2}} \\ 
			\displaystyle{-\dfrac{n^2}{2}} &\displaystyle{\xi s+\dfrac{n^2}{2}} \end{pmatrix} 
 \def\arraystretch{1}\left( \begin{array}{c}  \hat{\tilde{\rho}}_{+,1}\\ \hat{\tilde{\rho}}_{-,1} \end{array} \right)\\
 \Rightarrow X''=&MX\label{eq:Mdef}\,,
\end{align}
\end{subequations}
where we adopted the notation of \cit{balu2018role}: double primes indicate second partial derivatives on the vector $X=(\hat{\tilde{\rho}}_{+,1}, \hat{\tilde{\rho}}_{-,1})^{T}$.
We rewrite $M$ to $M=PDP^{-1}$ where
\begin{subequations}
\begin{align}
P&=\begin{pmatrix} \nu_{1}&\nu_{2} \\ 
			1 &1 \end{pmatrix} \,,
			&
D&=\begin{pmatrix} \mu^2&0 \\ 
			0 &\eta^2 \end{pmatrix} 	\,,		
\end{align}
\end{subequations}
with components given by
\begin{subequations}\label{eq:nurandm}
\begin{align}
\nu_{1} &=\frac{s(\xi-1)+\zeta}{n^2}\,, & \mu^2 &=\frac{1}{2}\left[n^{2}+s(1+\xi)-\zeta\right] \,,\label{eq:nurandm1}\\
\nu_{2} &=\frac{s(\xi-1)-\zeta}{n^2}\,, & \eta^2 &=\frac{1}{2}\left[n^{2}+s(1+\xi)+\zeta\right] \,,
\end{align}
\end{subequations}
with
\begin{equation}\label{eq:zeta}
\zeta=\sqrt{n^4+s^2(1-\xi)^2}\,.
\end{equation}
With $U=(u_{1}, u_{2})^{T}\equiv P^{-1}X$ we  rewrite \eqr{eq:Mdef} to  
 \mbox{$U''=DU$},
which is solved by $u_{1}=a_{1}\sinh \mu \tilde{x}$ and $u_{2}=a_{2}\sinh \eta \tilde{x}$,
with $a_{1}, a_{2}$ to be fixed by the boundary conditions. We return to our familiar  densities via $X=PU$, 
\begin{subequations}\label{eq:densities}
\begin{align}
\hat{\tilde{\rho}}_{+,1}&=\nu_{1}a_{1}\sinh \mu \tilde{x}  +\nu_{2}a_{2}\sinh \eta \tilde{x}\,,\\
\hat{\tilde{\rho}}_{-,1}&=a_{1}\sinh \mu \tilde{x}  +a_{2}\sinh \eta \tilde{x}\,.
\end{align}
\end{subequations}
Enforcing the Laplace-transformed b.c.'s [cf. \eqr{eq:electrokineticbcslowions}],
 \begin{align}
  \partial_{\tilde{x}}\hat{\tilde{\rho}}_{\pm,1}(\pm 1,\tilde{t}\,)&=-\frac{\alpha_{\pm}}{s}\,,
 \end{align}
yields 
\begin{subequations}
\begin{align}
\nu_{1}a_{1}\mu\cosh \mu  +\nu_{2}a_{2}\eta\cosh \eta &=-\frac{\alpha_{+}}{s}\,,\\
a_{1}\mu\cosh \mu  +a_{2}\eta\cosh \eta &=-\frac{\alpha_{-}}{s}\,,
\end{align}
\end{subequations}
for both boundaries (since $\cosh-x=\cosh x$). We 
 solve for $a_{1}$ and $a_{2}$,
\begin{align}\label{eq:a1a2}
a_{1}&=\frac{\alpha_{-}\nu_{2}-\alpha_{+}}{(\nu_{1}-\nu_{2})s\,\mu\cosh \mu}\,, \quad&
a_{2}&=\frac{\alpha_{-}-\alpha_{+}\nu_{1}}{(\nu_{1}-\nu_{2})s\,\eta\cosh \eta}\,,
\end{align}
and insert these results into \eqr{eq:densities} to find 
\begin{align}
\hat{\tilde{q}}_{1}(\tilde{x},s)&=\frac{n^2}{2\zeta s}\bigg[(\nu_{1}-1)(\alpha_{-}\nu_{2}-\alpha_{+})\frac{\sinh \mu \tilde{x}}{\mu\cosh \mu} \nn
&\quad\quad\quad\quad+(\nu_{2}-1)(\alpha_{+}-\alpha_{-}\nu_{1})\frac{\sinh \eta \tilde{x}}{\eta\cosh \eta}\bigg]\,.\label{eq:hatq}
\end{align}
In the special case $\xi=1$, Eqs.~\eqref{eq:nurandm} and \eqref{eq:zeta} reduce to $\nu_{1}=1$, $\nu_{2}=-1$, $\mu^2=s$, $\eta^2=n^2+s$, and $\zeta=n^2$,
and $\hat{\tilde{q}}_{1}$ reduces to \eqr{eq:laplaceq}.

Now, the following local electrostatic potential
\begin{align}\label{eq:localpotential}
\hat{\tilde{\psi}}_{1}(\tilde{x},s)&=\frac{n^4}{4\zeta }\bigg[(\alpha_{-}\nu_{2}-\alpha_{+})\frac{\nu_{1}-1}{s\, \mu^2}\left(x-\frac{\sinh \mu \tilde{x}}{\mu\cosh \mu}\right) \nn
&\quad\quad\quad+(\alpha_{+}-\alpha_{-}\nu_{1})\frac{\nu_{2}-1}{s \,\eta^2}\left(x-\frac{\sinh \eta \tilde{x}}{\eta\cosh \eta}\right)\bigg]\,,
\end{align}
satisfies both the Poisson equation \eqref{eq:Poissondimless} and one of its boundary conditions \eqr{eq:nofluxdimless}.
We do need to not enforce \eqr{eq:bcfloatingpsi}, as it trivially drops out of the thermovoltage, $\hat{V}_{T}(s)=\hat{\psi}(-1,s)-\hat{\psi}(1,s)$, 
the quantity of interest here. We find
\begin{subequations}\label{eq:thermovoltagelaplace}
\begin{align}
\hat{\tilde{V}}_{T}(s)&\equiv \hat{\tilde{V}}_{T}^{a}(s)+\hat{\tilde{V}}_{T}^{b}(s)+\mathcal{O}\left(\epsilon^2\right)\,,\\
\hat{\tilde{V}}_{T}^{a}(s)&=\frac{n^4 \epsilon}{2\zeta}(\alpha_{-}\nu_{2}-\alpha_{+})\frac{1-\nu_{1}}{s\,\mu^2}\left(1-\frac{\tanh \mu}{\mu}\right)\,,\\
\hat{\tilde{V}}_{T}^{b}(s)&=\frac{n^4 \epsilon}{2\zeta}(\alpha_{+}-\alpha_{-}\nu_{1})\frac{1-\nu_{2}}{s\,\eta^2}\left(1-\frac{\tanh \eta}{\eta}\right)\,.
\end{align}
\end{subequations}
Besides the pole at $s=0$, which determines the steady state of $V_{T}$, the poles of $\hat{\tilde{V}}_{T}(s)$ with nonzero residues 
appear in the $\tanh \mu$  and $\tanh \eta$ terms of \eqr{eq:thermovoltagelaplace}, and lie at $\mu =\pm i \mathcal{N}_{j} $ and $\eta=\pm i \mathcal{N}_{j}$.
With \eqr{eq:nurandm1}  we write  
\begin{align}
&n^{2}+s(1+\xi)-\sqrt{n^4+s^2(1-\xi)^2} =-2\mathcal{N}_{j}^2\,,
\end{align}
which has two solutions for each $j$:
\begin{align}
s^{j}_{\pm}&=\mp\frac{1}{4\xi}\sqrt{n^4(1+\xi)^2+4n^2\mathcal{N}_{j}^2(1-\xi)^2+4\mathcal{N}_{j}^4(1-\xi)^2}\nn
&\quad-\frac{1}{4\xi}\left(n^2+2\mathcal{N}_{j}^2\right)(1+\xi)\,.\label{eq:ssolutions1}
\end{align}
As we are interested in $n\gg1$, we report
\begin{align}
s^{j}_{\pm}\overset{n\gg1}=&\mp \frac{1}{4\xi} n^2(1+\xi)\left(1+2\frac{\mathcal{N}_{j}^2}{n^2}\frac{(1-\xi)^2}{(1+\xi)^2}\right)\nn
&-\frac{1}{4\xi}\left(n^2+2\mathcal{N}_{j}^2\right)(1+\xi)+\mathcal{O}(n^{-2})\,.\label{eq:ssolutions2}
\end{align}
This amounts to
\begin{align}
s^{j}_{+}&=-\frac{n^2(1+\xi)}{2\xi}-\mathcal{N}_{j}^2\frac{1+\xi^2}{\xi(1+\xi)}+\mathcal{O}(n^{-2})\,,\nn
s^{j}_{-}&=-\frac{2\mathcal{N}_{j}^2}{1+\xi}+\mathcal{O}(n^{-2})\,.\label{eq:sresults}
\end{align}
Interestingly, $\cosh \eta=0$ has the same $s^{j}_{\pm}$ solutions. 
To determine which $s^{j}_{\pm}$ solutions are physically relevant, we take $\xi=1$ in \eqr{eq:ssolutions1} and find
$s^{j}_{\pm}=-n^2(1\pm1)/2-\mathcal{N}_{j}^2$.
Hence, we retrieve the $\xi=1$ timescales if we take $s^{j}_{-}$ for the $\cosh \mu=0$ poles, and $s^{j}_{+}$ for the $\cosh \eta=0$ poles.
 
The $n\gg1$ behavior of Eqs.~\eqref{eq:nurandm} and \eqref{eq:zeta} evaluated at $s^{j}_{-}$ reads 
\begin{subequations}\label{eq:nuandrsm}
\begin{align} 
\zeta(s^{j}_{-})&=n^2+\mathcal{O}(n^{-2})\,,\\
\nu_{1} &=1-\frac{s^{j}_{-}(1-\xi)}{n^2}+\mathcal{O}(n^{-4})\,,\\
\nu_{2} &=-1-\frac{s^{j}_{-}(1-\xi)}{n^2}+\mathcal{O}(n^{-4})\,,\\
\mu^2 &=\frac{s^{j}_{-}(1+\xi)}{2}+\mathcal{O}(n^{-2})\,,
\end{align}
\end{subequations}
while at $s^{j}_{+}$ we find
\begin{subequations}\label{eq:nuandrsq}
\begin{align}
\zeta(s^{j}_{+})&=\frac{n^2(1+\xi^2)}{2\xi}
+\frac{\mathcal{N}_{j}^2(1-\xi)^2}{\xi}+\mathcal{O}(n^{-0})\,,\\
\nu_{1} &=\frac{1}{\xi}+\mathcal{O}(n^{-2})\,, \\
\nu_{2} &=-\xi+\mathcal{O}(n^{-2})\,,\\
\eta^2 &=-\mathcal{N}_{j}^2+\mathcal{O}(n^{-2})\,.
\end{align}
\end{subequations}
Similar to \eqr{eq:A6} we find
\begin{subequations}\label{eq:coshclosetosm}
\begin{align}
\frac{\tanh \mu}{\mu}\overset{s\to s^{j}_{-}}= &\frac{4}{1+\xi}\frac{1}{s-s^{j}_{-}}\label{eq:coshclosetosm1}\,,\\
\frac{\tanh \eta}{\eta}\overset{s\to s^{j}_{+}}{=} &\frac{4}{1+\xi}\frac{1}{s-s^{j}_{+}}\label{eq:coshclosetosm2}\,.
\end{align}
\end{subequations}
Inserting Eqs.~\eqref{eq:nuandrsm} and \eqref{eq:coshclosetosm1} into $\hat{V}_{T}^{a}(s) $ gives 
\begin{align}\label{eq:thermovoltagelaplace2}
\hat{\tilde{V}}_{T}^{a}(s)
\overset{s\to s^{j}_{-}}{\sim}-2\epsilon\alpha_{\rm s}\frac{1-\xi}{1+\xi}\frac{1}{\mathcal{N}_{j}^2(s-s^{j}_{-})}+\mathcal{O}(n^{-2})\,,
\end{align}
while inserting Eqs.~\eqref{eq:nuandrsq} and \eqref{eq:coshclosetosm2} into $\hat{V}_{T}^{b}(s) $ gives 
\begin{align}\label{eq:thermovoltagelaplace3}
\hat{\tilde{V}}^{b}_{T}(s)&\overset{s\to s^{j}_{+}}{\sim}-4\epsilon\frac{\xi\alpha_{+}-\alpha_{-}}{1+\xi}\frac{1}{\mathcal{N}_{j}^2(s-s^{j}_{+})}+\mathcal{O}(n^{-2})\,.
\end{align}  
We used proportionality signs in Eqs.~\eqref{eq:thermovoltagelaplace2} and \eqref{eq:thermovoltagelaplace3} because we disregarded the 1's in the bracketed terms of \eqr{eq:thermovoltagelaplace}, as their residues are zero at $s^{j}_{-}$ and $s^{j}_{+}$, respectively.
Calculating $V_{T}(t)=\sum_{s_{\xi}}\textrm{Res}\left(\hat{V}_{T}(s)\exp{(st)},s_{\xi}\right)$, with $s_{\xi}=\{0,s^{j}_{-},s^{j}_{+}\}$, now gives \eqr{eq:VTfinal}.

\end{appendix}

\end{document}